\documentclass[%
 amsmath,amssymb,
 aps,
]{revtex4-1}

\usepackage{graphicx}
\usepackage{dcolumn}
\usepackage{bm}
\usepackage{amsfonts}
\usepackage{yfonts}
\usepackage{amsmath}
\usepackage{amssymb}
\usepackage{comment}
\usepackage{epsfig}
\usepackage{float} 
\usepackage{hyperref}

\usepackage{color}

\begin{document}

\preprint{APS/123-QED}

\title{  ``Magneto-elastic'' waves in an anisotropic  magnetised plasma}

\author{{D. Del Sarto}}
 \affiliation{  Institut Jean
Lamour, UMR 7198 CNRS - Universit\'e de Lorraine, BP 239 F-54506 Vandoeuvre-le-Nancy, France}
 \email{daniele.del-sarto@universite-lorraine.fr}
\author{{F. Pegoraro}}%
\affiliation{Dipartimento di Fisica ``E. Fermi'', Universit\`a di Pisa and CNISM, Largo B. Pontecorvo 3, 56127 Pisa,  Italy}%
\author{A. Tenerani}%
\affiliation{EPSS, UCLA, Los Angeles, CA, USA }

\date{\today}
\begin{abstract}

 The  linear waves that propagate in a two fluid magnetised plasma allowing for a non-gyrotropic perturbed ion pressure tensor  are investigated.  For perpendicular propagation and perturbed fluid velocity a low frequency  (magnetosonic) and a high frequency (ion Bernstein) branch are identified and discussed.  For both branches a comparison is made with  
the results of a truncated Vlasov  treatment.  For the low frequency branch we show that a consistent expansion procedure  allows us to recover the correct expression of the {Finite Larmor Radius} corrections to the magnetosonic dispersion relation.

\end{abstract}

\maketitle


\section{Introduction}\label{Intro}

The inclusion of the full pressure tensor evolution in the  fluid equations  of a magnetised plasma  allows for a self-consistent description  within a fluid framework of important effects that would otherwise require a kinetic treatment. {This approach was  used, e.g.  to study collisionless magnetic reconnection driven by electron pressure anisotropy  \cite{Hesse,Kutznetsova1,Kutznetsova2,Yin1,Yin2} or to
investigate the dispersion relation of Geodesic Acoustic Modes \cite{Smol1,Smol2}, within the so-called Grad hydrodynamic equations framework recovering in this way ``kinetic'' effects  at a much lower computational cost.
In addition, a macroscopic point of view can help  to  evidentiate dynamical processes that may be difficult to identify within the more complex kinetic description.} 
In particular  {it was recently shown that} a spatial inhomogeneity in  the fluid velocity shear can make  the pressure tensor  non-gyrotropic \cite{Tensor_Anis}: in the case of a  spatially localised sheared flow, the formation   of a non-gyrotropic ion pressure tensor was shown to be mediated  by the propagation in the direction orthogonal to the magnetic field of  high frequency  waves  that can be interpreted as the fluid counterpart of an ion  Bernstein branch.
\,
Thus it is  of interest to examine in  detail these extended fluid models in which a full pressure tensor dynamics is included and to evidentiate their 
features together with their limitations. With this aim, here we  restrict our analysis to the case of an ion pressure tensor and investigate the small amplitude limit of the plasma description adopted in Ref.\cite{Tensor_Anis}. More specifically we
derive the  dispersion relation of the linear waves, propagating in the plane perpendicular to a strong magnetic field, described by this model and compare them to the  corresponding kinetic results obtained by an appropriate expansion of the solution of  the linearised Vlasov-Maxwell system. 
We  call these waves ``Magneto-elastic'' waves.

The possibility of treating ion Bernstein modes within a fluid formalism, possibly extending the fluid equations to include high {order} velocity moments of the  particle distribution function, may provide a convenient investigation  tool of plasma dynamics far from thermodynamic equilibrium:  for example  it
 has been shown within a full Vlasov-Maxwell modelling that the excitation of $\omega \sim 4\Omega$  ion Bernstein waves (where $\Omega$ is the ion-cyclotron frequency) in the same geometry configuration we  consider here induces a spatially asymmetric, non-gyrotropic heating of ions \cite{Chiara1,Chiara2}.  In a different context ion Bernstein modes are of interest  in dedicated magnetically confined plasma experiments on ion-cyclotron heating mechanisms, such as, e.g., the ALINE \cite{ALINE} and SSWICH \cite{SSWICH} experiments that  aim to shed light on the convective acceleration and heating of particles in the radio-frequency sheets, which develop  both next to the tokamak walls because of the radio-frequency discharges and  next   to antennas for the ion-cyclotron heating. In Ref.\cite{Amatucci}  it has been experimentally demonstrated  that sheared plasma flows which result from inhomogeneous transverse electric fields in a magnetised plasma induce the excitation of waves  propagating in the ion-cyclotron frequency range almost perpendicularly to the magnetic field. Conversely, the  excitation of Ion-Bernstein waves is known, both experimentally \cite{LeBlanc} and theoretically (see Ref.\cite{Myra} and references therein),  to induce the generation of sheared ion flows.

Non-gyrotropic particle distribution functions are measured  in the solar wind and at the solar wind-magnetosphere interface and correlation with the presence of shear flows has  been evidenced.  For example, non-gyrotropic distribution functions have been measured within a {magnetic flux tube of ions flowing out} into the upstream solar wind \cite{Posner}. Non-gyrotropic electron distributions have been observed within the electron diffusion region of reconnecting magnetic structures in the magnetopause \cite{Scudder_1,Scudder_2} near  $X$- and  $O$-points,  where  the {stream function of the} velocity field is locally hyperbolic (see e.g. Ref.\cite{PPCF}).

In the following we  consider the propagation of ``Magneto-elastic'' waves   in  
a  homogeneous Double Adiabatic (Chew-Goldberger-Low, or CGL)  plasma equilibrium embedded in an externally imposed magnetic field and look for perturbations with  a wave-vector perpendicular to the magnetic field.  
For the sake of simplicity we take cold, massless electrons.  We find two branches  of perturbations with  fluid velocities and electric fields polarised  in the plane perpendicular to the magnetic field:  a low frequency branch (LFB) corresponding to magnetosonic waves and a high frequency branch (HFB) corresponding to (generalized i.e., non quasi-electrostatic) ion Bernstein modes. 
We analyse their dispersion relations and compare their long wave-length  behaviour   to the kinetic counterparts obtained by linearizing the corresponding Vlasov-Maxwell system  and by retaining the $n=0,\, \pm 1,\,\pm 2$ ion cyclotron harmonics in the small Larmor-radius limit. 
For the sake of brevity in the following we will  refer  to the  dispersion relations obtained from  such a   truncated kinetic Vlasov  ion response as the ``kinetic'' dispersion relations.
 Qualitative and quantitative differences in the respective dispersion relations are evidenced. For the low frequency branch  we discuss  a discrepancy present in the literature on the derivation of the Finite Larmor Radius (FLR) corrections  in the  CGL-FLR fluid limit.

This article begins with the description of the fluid model (Sec.\ref{Fluid}) and with its linear analysis in the  case of 1D perturbations propagating perpendicularly to a background magnetic field (Sec.\ref{Linear_fluid}). In Sec.\ref{Vlasov} the internal consistency  of the model is discussed and its limitations are identified by direct comparison with the corresponding  kinetic  dispersion
 relations. 
In Sec.\ref{Conclusion} a summary of the results obtained is presented {and the possibility of describing higher ion cyclotron harmonics ($|n|> 1$) within  a fluid framework is noted.}  
 Detailed calculations are presented in the Appendices: in Appendix  \ref{Appendix_A}  a brief derivation of the two-fluid model equations is given, in Appendix \ref{Appendix_B} the limit of the Double Adiabatic closure is described and in Appendix \ref{Appendix_C}  the derivation of the FLR limit of the magnetosonic branch is detailed. 
 We recall in this context  that  the Double Adiabatic  closure  (see e.g. \cite{Krall})  describes the low frequency, long wavelength dynamics of a magnetized plasma with different temperatures, and obeying different equations of state,  in the direction of the magnetic field and in the plane perpendicular to it.
It assumes temperature isotropy 
in this plane. This  closure  applies to  plasma conditions where heating and cooling processes   and/or approximate integrals of motion  due to the presence of particle adiabatic invariants affect the parallel and the perpendicular directions differently while it assumes 
that the smallness of the particle gyroradii and their fast gyration are sufficient to allow for a fluid isotropic description in the perpendicular plane. 
For the perpendicular propagation considered here   only  fast magnetosonic waves  could  propagate if  the Double Adiabatic description were applied to the perturbations of the equilibrium.

\section{Fluid model with  full pressure tensor evolution}\label{Fluid}
We consider a  single-fluid MHD model in which the contributions of the electron pressure  and of the electron inertia are  neglected and the ion pressure  evolves according to the full pressure tensor equation.
This model can be obtained from the  two-fluid, full pressure tensor  equations derived in Appendix \ref{Appendix_A}  by adding in the standard way the  ion  and electron momentum equations (Eq.(\ref{eq:A_3})) and by using the quasi-neutrality assumption $n^e \sim n^i$ for a hydrogen plasma. 
Thus the magnetic force  and  the spatial derivatives of the ion pressure tensor  are the only forces acting on the  plasma (Eq.(\ref{eq:M_2})), while the electron momentum equation appears in the single-fluid system in the form of the Hall-MHD Ohm's law in the induction equation (Eq.(\ref{eq:M_5})). 
Furthermore   we disregard  the ion heat flux, $\partial_i Q_{ijk}=0$  in the   ion  pressure tensor equation (Eq.(\ref{eq:A_6})).  In the present  analysis  this simplifying assumption is made   as a closure condition whose consistency with the truncated Vlasov model in the low-$\beta$ limit will be shown later (Sec.\ref{Vlasov}). Here $\beta$ is the  ratio (suitably defined, see Eq.(\ref{eq:L_fluid_3})) between the ion pressure and the magnetic pressure.  We  note however that, as stressed in Ref.\cite{Mikhailovskii}, the comparison  with the kinetic dispersion relation may be made  exact  in the context of  the low frequency limit of  magneto-sonic waves also at $\beta \sim 1$ if the heat-flux contribution is consistently retained (see Sec.\ref{beta} and Appendix \ref{Appendix_C}).  \\
Introducing  the ion cyclotron frequency $\Omega$ and the unit vector components $b_i\equiv B_i/B$, the  single-fluid equations  read, with standard  notation,
\begin{equation}\label{eq:M_1}
\partial_t n\,
\,+\,{\bm\nabla}\cdot(n{\bm u})\,=\,0,
\end{equation}
\begin{equation}\label{eq:M_2}
\partial_t {\bm u}\,+\,{\bm u}\cdot{\bm \nabla}{\bm u}\,=
\Omega  \, \frac{\bm J}{n e} \times{\bm b}\,
-\,\frac{ {\bm\nabla} \cdot {\bm\Pi} }{mn},
\end{equation}
\begin{equation}\label{eq:M_3}
\partial_t{\bm \Pi}
 \,+\,{\bm\nabla}\cdot({\bm u}\,{\bm \Pi})
\,+\, {\bm \Pi}\cdot{\bm\nabla}{\bm u}
\,+\, ({\bm \Pi}\cdot{\bm\nabla}{\bm u})^{T}
 \end{equation}
$$\,-\,\Omega
 ({\bm \Pi}\times{\bm b} + ({\bm \Pi}\times{\bm b})^{T} )\,=\,0.$$ 
The apex ``$T$'' indicates   matrix  transpose. Note in passing that Eq.(\ref{eq:M_3}) ensures that the pressure tensor remains positive definite over time if it is positive definite at $t=0$.\\
 In the adopted model the displacement current is neglected in Amp\`ere's law (Eqs.(\ref{eq:A_7})), as consistent with the quasi-neutrality assumption and the restriction to phase velocities  smaller that the speed of light  thus leading to the MHD definition of the current density, 
\begin{equation}\label{eq:M_4}
{\bm J}=\frac{c}{4\pi} {\bm \nabla}\times{\bm B}.
\end{equation}
Then the induction equation coupled with Ohm's law is 
\begin{equation}\label{eq:M_5}
\partial_t {\bm B}={\bm \nabla}
\times\left\{\left({\bm u}-\frac{\bm J}{ne}\right)\times{\bm B} \right\}.
\end{equation}
The third and  the fourth terms  in Eq.(\ref{eq:M_3}) arise from  the action  on the pressure tensor ${\bm\Pi}$ of the   spatial derivatives of the velocity field ${\bm u}$  which we write in tensor notation as the sum of  a strain (that is, of compression and of compressionless shear)  and of a vorticity term as 

{\begin{equation}\label{eq:M_6}
\frac{\partial u_i}{\partial x_j}
 =\underbrace{\underbrace{
\frac{1}{3}\left(
{\frac{\partial u_k}{\partial x_k}}\right)\delta_{ij}
}_\text{compression rate}\,
+\underbrace{\frac{1}{2}\left[\left(\frac{\partial u_i}{\partial x_j}
+\frac{\partial u_j}{\partial x_i}\right)
-\frac{2}{3}\left(
\frac{\partial u_k}{\partial x_k}\right)\delta_{ij}
\right]}_{\text{compressionless rate of shear}}}_{\text{strain tensor}}
+\,\underbrace{\frac{1}{2}\left(\frac{\partial u_i}{\partial x_j}
- \frac{\partial u_j}{\partial x_i}\right)}
_{\text{fluid vorticity}} .
\end{equation}}

\section{Linear analysis of the fluid model}\label{Linear_fluid}
We linearise Eqs.(\ref{eq:M_1}-\ref{eq:M_5}) around a  double adiabatic equilibrium $\Pi_{ij}^0=P^0_\perp\delta_{ij}+(P^0_{||}-P^0_\perp)b_ib_j$ (see App.\ref{Appendix_B})  with uniform density $n_0$, no velocity (${\bm u}^0=0$), uniform magnetic field ${\bm B}^0=B_0{\bm e}_z$. A double adiabatic equilibrium is the most general,  spatially uniform configuration  compatible with  the  pressure tensor  equation (Eqs.(\ref{eq:C_1}-\ref{eq:C_2})) in the absence of a  shear flow \cite{CGL}.  When the latter is included, a class of spatially inhomogeneous, non-gyrotropic equilibria can be found  \cite{Cerri_2013,Cerri_2014}.  Here we  consider  perturbations with  a wave-vector perpendicular to the equilibrium magnetic field, ${\bm k}=k{\bm e}_x$, since this makes it possible  to decouple the dispersion relation of  modes with electric and velocity field components perpendicular  to the equilibrium magnetic field $B_0{\bm e}_z$  from the modes with parallel components. This simplifies the  study of the propagation of perturbations that make  the pressure non-gyrotropic  in the  plane perpendicular to  $B_0{\bm e}_z$. Note however that  the curl of the Hall term in Eq (\ref{eq:M_5}) vanishes in this geometry.\\
 Labelling   perturbed quantities by a  \;$\tilde{  }$\; we  set  \begin{equation}\label{eq:L_fluid_0}
\tilde{\bm u} = \tilde{\bm u}_\perp e^{i(kx - \omega t)} + \tilde{u}_z {\bm e}_z e^{i(kx - \omega t)}.
\end{equation}
 It is convenient to introduce the characteristic Alfv\'en velocity ($c_a$) and the ion thermal velocity ($v_{th}$) as
\begin{equation}\label{eq:L_fluid_1}
c_a^2\equiv\frac{B_0^2}{4\pi n_0m},\qquad\qquad  v_{th}^2\equiv 2\frac{k_BT}{m}= 2\frac{P^0_\perp}{n_0 m}.
\end{equation}
We then introduce the ion skin depth ($d_i$) and the thermal ion Larmor radius ($\rho_i$),
\begin{equation}\label{eq:L_fluid_2}
d_i^2=\frac{c_a^2}{\Omega^2},\qquad\qquad \rho_i^2=\frac{v_{th}^2}{\Omega^2}.
\end{equation}
In this model in which only the ion  temperature contributes to the plasma pressure, the scale separation between $d_i$ and $\rho_i$ is measured by the magnitude of the $\beta$ parameter,

\begin{equation}\label{eq:L_fluid_3}
\beta\equiv\frac{v_{th}^2}{c_a^2}=\frac{\rho_i^2}{d_i^2}.
\end{equation}

Linearising Eq.(\ref{eq:M_2}) in this geometry we find the equations for the in-plane velocity components,
\begin{equation}\label{eq:linear_ux}
\tilde{u}_x \;= \; \left(\frac{k}{\omega}\right)^2c_a^2\tilde{u}_x 
+ \,\frac{k}{\omega}\frac{\tilde{\Pi}_{xx}}{n_0m},
\end{equation}
\begin{equation}\label{eq:linear_uy}
\tilde{u}_y \; = \; \frac{k}{\omega}\frac{\tilde{\Pi}_{xy}}{n_0m},
\end{equation}
which are coupled to the equations for the perpendicular pressure components, as obtained from Eqs.(\ref{eq:M_3}), 
 \begin{equation}\label{eq:linear_tens_xx}
\frac{\omega}{\Omega}\tilde{\Pi}_{xx}\; =\; 3P^0_{\perp}
\frac{k\tilde{u}_x}{\Omega} + 2i\tilde{\Pi}_{xy},
\end{equation}
\begin{equation}\label{eq:linear_tens_xy}
\frac{\omega}{\Omega}
\tilde{\Pi}_{xy}\; =\; P^0_{\perp}
\frac{k\tilde{u}_y}{\Omega} +i(\tilde{\Pi}_{yy}-\tilde{\Pi}_{xx}),
\end{equation}
\begin{equation}\label{eq:linear_tens_yy}
\frac{\omega}{\Omega}
\tilde{\Pi}_{yy}\; =\; P^0_{\perp}
\frac{k\tilde{u}_x}{\Omega} - 2i\tilde{\Pi}_{xy},
\end{equation}
{and to the equation for} the parallel pressure component. {The latter evolves}  because of the plasma compressibility,
\begin{equation}\label{eq:linear_tens_zz}
{\omega}
\tilde{\Pi}_{zz}\; =\; P^0_{||}k\tilde{u}_x.  
\end{equation}
After a few algebraic steps the linearised system equations can be cast in the form $[{\bm M}]\cdot\tilde{\bm u}=0$, with the dispersion matrix given by:
\begin{equation}\label{eq:L_fluid_3}
[{\bm M}]\equiv\left(
\begin{array}{cc}
\displaystyle{1-\frac{k^2(c_a^2 + v_{th}^2)}{\omega^2}
+\frac{k^2v_{th}^2}{2(4\Omega^2-\omega^2)}}
 & \qquad 
\displaystyle{i\frac{\Omega}{\omega}
\frac{k^2v_{th}^2}{(4\Omega^2-\omega^2)}} 
  \\ \\
-\displaystyle{i\frac{\Omega}{\omega}
\frac{k^2v_{th}^2}{(4\Omega^2-\omega^2)}}
& 
 \displaystyle{1
+\frac{k^2v_{th}^2}{2(4\Omega^2-\omega^2)}} 
\\ \\
\end{array}
\right).
\end{equation}

A further set of closed equations is provided by the parallel component of the velocity, $\tilde{u}_z$, which is decoupled from the perturbations $(\tilde{u}_x, \tilde{u}_y,0)$ but is coupled to the linear equations for $\tilde{\Pi}_{xz}$ and $\tilde{\Pi}_{yz}$,
\begin{equation}\label{eq:linear_uz}
\tilde{u}_z \; =  \; \frac{k}{\omega}\frac{\tilde{\Pi}_{xz}}{n_0m},
\end{equation}
\begin{equation}\label{eq:linear_tens_xz}
\frac{\omega}{\Omega}
\tilde{\Pi}_{xz}\; =\; P^0_{\perp}
\frac{k\tilde{u}_z}{\Omega} + i\tilde{\Pi}_{yz},
\end{equation}
\begin{equation}\label{eq:linear_tens_yz}
\frac{\omega}{\Omega}
\tilde{\Pi}_{yz}\; =\; - i\tilde{\Pi}_{xz}.
\end{equation}
This set provides the dispersion relation of a spurious mode which is unphysical having no counterpart in the Vlasov description:
\begin{equation} \label{eq:ugly_duckling}
\left( 1
+\frac{k^2v_{th}^2}{2(\Omega^2-\omega^2)}\right) \tilde{u}_z=0,
\end{equation}  
(the differences with respect to the corresponding branch obtained from a kinetic treatment are evidenced in Sec.\ref{Vlasov}).
 {In fact this mode  would correspond to a hydrodynamic mode
 with no perturbed electromagnetic fields and thus cannot be realized in a collisionless plasma.  
 This feature   originates from  the simplified description of the parallel dynamics adopted in the present  fluid model which,  for ${\bm k}\perp {\bm B}_0$ and in  the limit of cold mass-less electrons,  does not make it possible to  balance a  parallel electric field component but it 
does not affect the  modes  with perturbed velocity perpendicular to the equilibrium magnetic field treated in the rest of this paper.}
  {We  call the modes  described by Eqs.(\ref{eq:linear_ux}-\ref{eq:linear_tens_zz})  ``\emph{magneto-elastic} modes'', as they arise because of the  pressure anisotropy which couples the electromagnetic perturbations described by Eq.(\ref{eq:linear_ux}) to elastic-type perturbations related to the $\tilde{u}_y$ component (Eq.(\ref{eq:linear_uy})) through the off-diagonal component  $\tilde{\Pi}_{xy}$.}
 The evolution of $\Pi_{zz}$, though not directly contributing to the dynamics of $\tilde{\bm u}_\perp$, is coupled to it (see Eq.(\ref{eq:linear_tens_zz})) because of the compression related to the velocity component $\tilde{u}_x$ (Sec.\ref{B_z}). \\
{When   comparing the  full pressure tensor dispersion relation with the isotropic MHD model  where the magnetosonic branch has dispersion relation $\omega^2=k^2(c_{A}^2+c_s^2)$, with $c_s$ sound speed of the plasma, we need to recall   that,  since  the electrons  are taken to be cold,  here the sound velocity is replaced by $v_{th}$ of Eq.(\ref{eq:L_fluid_1}). In particular, the factor $2$ in the definition of $v_{th}^2$   corresponds to a polytropic index $\Gamma_\perp=2$, as follows from the double adiabatic equation for ${P_\perp}$ (Eq.\ref{eq:C_6}), when ${\bm B}_0\cdot{\bm k}=0$. }

\begin{figure}
\centerline{\epsfig{figure=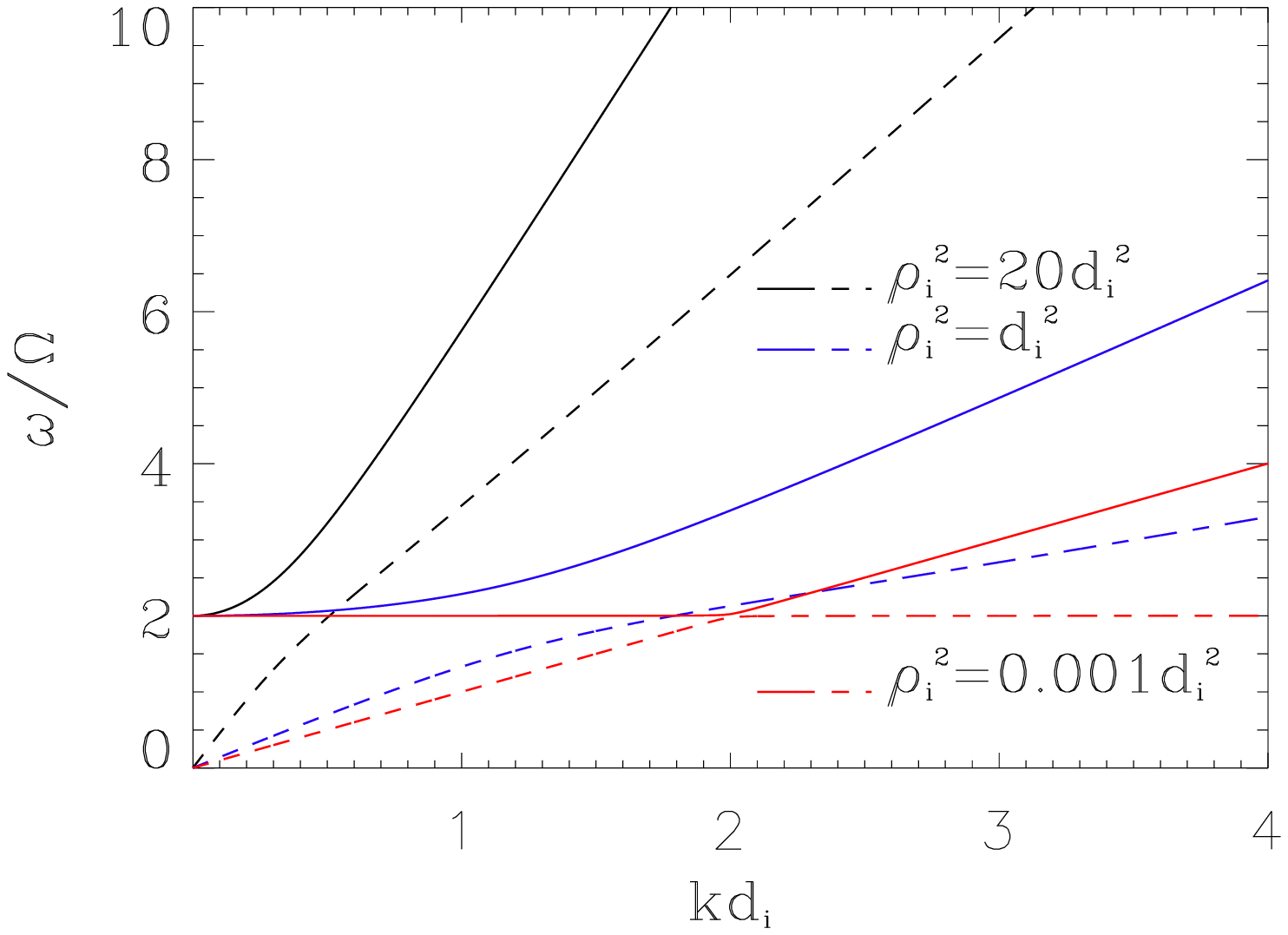,width=8.cm}
\epsfig{figure=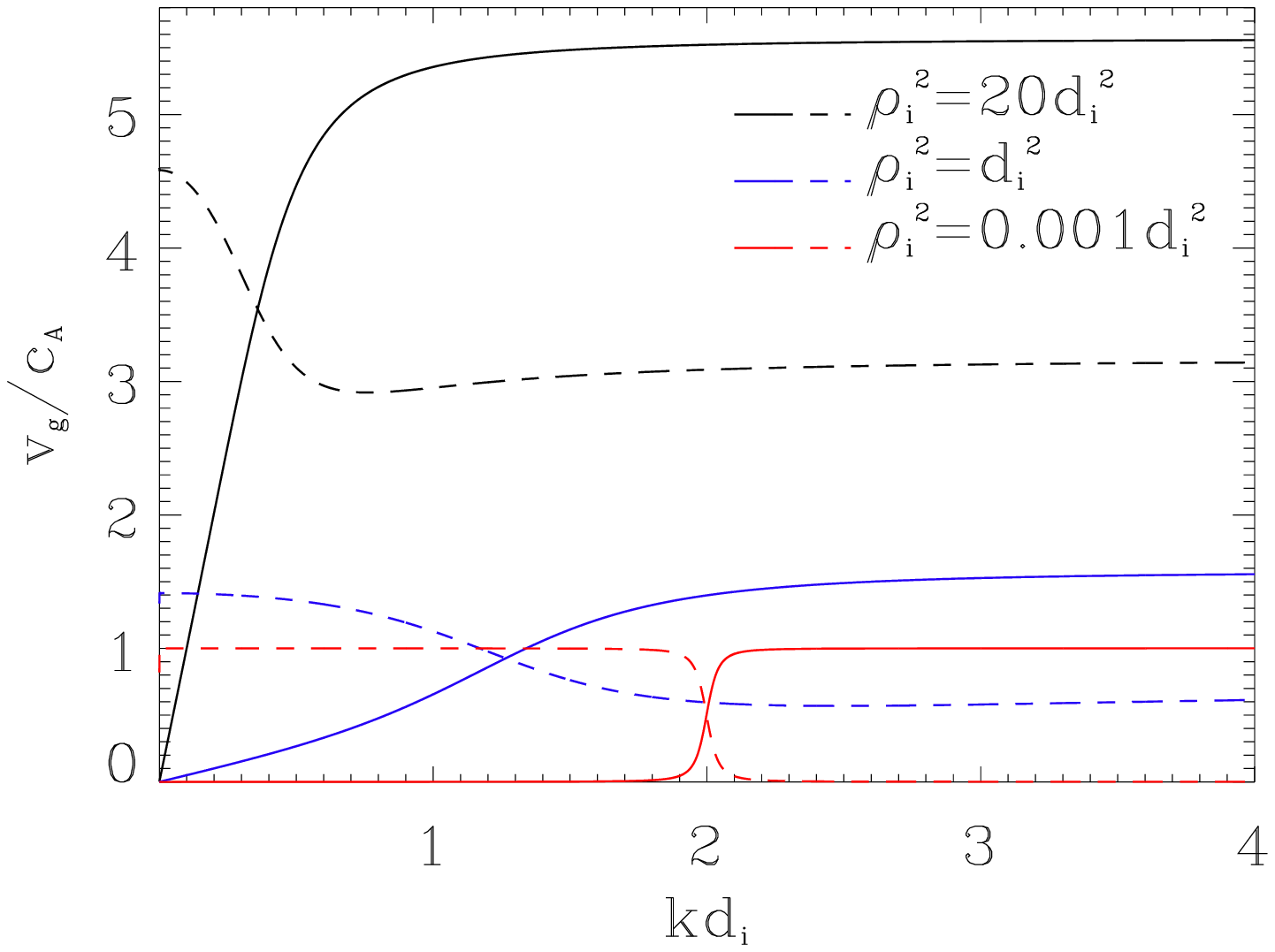,width=8.cm}}
\caption{Dispersion relations (left frame) and
group velocities   versus $k d_i$   (right frame) of the  LFB ($\omega_l$ and $v_l$, dashed lines)  and of  the HFB ($\omega_h$ and $v_h$, solid lines) obtained from the fluid model for different values of $v_{th}/c_{A}=\rho_s/d_i$.  }\label{Fig_disp_tensor-0}
\end{figure}

 It is  convenient to write  the dispersion relation of the modes $\tilde{\bm u}=\tilde{\bm u}_\perp$ as
\begin{equation}\label{eq:L_fluid_4}
\frac{\displaystyle{\left(\frac{\omega^2}{\Omega^2}-4-\frac{k^2\rho_i^2}{2}\right)
\left[\frac{\omega^2}{\Omega^2} -k^2\left(d_i^2+\frac{3}{2}\rho_i^2\right)\right]
-2k^2\rho_i^2}}
{\omega^2(4\Omega^2-\omega)}=0,
\end{equation}
which describes the two branches
\begin{equation}\label{eq:L_fluid_5}
\left(\frac{\omega_{h,l}}{\Omega}\right)^2 = 2 + k^2\left(\frac{d_i^2}{2}+\rho_i^2 \right)
 \pm 2 \sqrt{ \left( 1 - \frac{k^2}{4}\left(d_i^2+\rho_i^2\right)\right)^2 +\frac{k^2\rho_i^2}{2} }, 
\end{equation}
where,  consistent with the notation of  Ref.\cite{Tensor_Anis}, the higher frequency  branch (HFB), corresponding to the root with the  plus sign in front of the square root,   is denoted by $\omega_h$ while the lower frequency branch (LFB), corresponding to the minus sign, is denoted by $\omega_l$.  The  behaviour  of these dispersion relations is exemplified in Fig.\ref{Fig_disp_tensor-0}, left frame. For $\omega  <2\Omega$ only the LFB propagates. It corresponds to a fast magnetosonic wave, corrected by the presence of the full pressure tensor evolution.  From a local expansion of the matrix $[{\bm M}]$ we see that  this branch is  compressive and has a  purely transverse  electric field at low frequencies {($\omega/\Omega\ll 1$)}, with eigenmodes of components
 \begin{equation}\label{eq:L_polarization_LFB}
{\left(\tilde{u}_x,\tilde{u}_y\right)
=\left(1,\,i\,\frac{\Omega}{\omega_l}\frac{k^2\rho_i^2}{4} \right),\qquad
(\tilde{E}_x,\tilde{E}_y)=
\left(-i\,\frac{\Omega}{\omega_l}\frac{k^2\rho_i^2}{4},\,1\right),}
\end{equation}
 while it becomes approximately right-hand circularly polarised, $(\tilde{u}_x,\tilde{u}_y)\sim i (\tilde{E}_x,\tilde{E}_y)
{\simeq (1,i)}$ at  the crossing 
of the ``resonance''  $\omega\simeq  2\Omega$.
At $\omega= 2\Omega$ the HFB appears  at $k=0$: its  dependence on $k$ is initially flat and  grows linearly  for larger values of $k$. This HFB is related to the $n=2$  ``generalised'' (i.e. not quasi-electrostatic \cite{Fredricks,Fredricks2}) ion-Bernstein mode which is found in a kinetic description (see Sec.\ref{Comparison}).  For $kd_i\ll 1$ this mode is left-hand circularly polarised with $(\tilde{u}_x,\tilde{u}_y)\sim i(\tilde{E}_x,\tilde{E}_y)\simeq (1, -i)$, and  maintains approximately  the same polarisation for $kd_i\sim 1$ (as deduced by a local expansion of the eigenmode equations around the resonance).  This branch is not present in a double adiabatic description and it arises within a fluid description because the full  pressure tensor  equations allow for non-gyrotropic  pressure perturbations  in the $x$-$y$ plane.  The lower frequency bound of this branch,  $\omega = 2\Omega$,  follows from the fact that, because  of the equilibrium  magnetic field,  these components rotate in the $x$-$y$ plane  at twice the cyclotron frequency. \\ The group velocity of the two branches, normalised to the Alfv\'en velocity, can be expressed as 
\begin{equation}\label{eq:L_fluid_7}
\frac{v_{h,l}}{c_a} = \frac{kc_a}{\omega_{h,l}}
\left\{\displaystyle{ \frac{1}{2}+\frac{\rho_i^2}{d_i^2}
\pm \frac{A}{B} }\right\}, 
\end{equation}
where  the $\pm$ sign in front of last term corresponds to the
 high and to the low branches respectively and
\begin{equation}\label{eq:L_fluid_8} 
A = -1+k^2\left(\frac{1}{4}+
 \frac{\rho_i^2}{2 d_i^2} \right) + \frac{k^2\rho_i^4}{4 d_i^2},\end{equation}
\begin{equation}\label{eq:L_fluid_9} 
B =2 \sqrt{ \left( 1 - \frac{k^2}{4}\left(d_i^2+\rho_i^2\right)\right)^2 +\frac{k^2\rho_i^2}{2}}.
\end{equation}
Its  behaviour  is exemplified in Fig.\ref{Fig_disp_tensor-0}, right frame, for some values of $d_i$ and $\rho_i $.

  Note  that  in the limit $\rho_i \to 0$ (cold ion limit, vanishing pressure tensor)  the two branches  come close to crossing  each other  (see left frame of Fig.\ref{Fig_disp_tensor-0} ) and that  for  $\rho_i = 0$ the HFB disappears  (in formal terms the solution $\omega^2/\Omega^2 - 4 =0$ of Eq.(\ref{eq:L_fluid_5}) is cancelled by  the corresponding resonant term in the denominator of Eq.(\ref{eq:L_fluid_4})). \\ Ordering $\rho_i\sim d_i$, in the long  wavelength limit $(\rho_i^2 \sim d_i^2, \; k\rightarrow 0)$ we obtain
{\begin{equation}\label{eq:L_fluid_limit1} 
\omega_l^2 \simeq k^2(c_a^2+v_{th}^2),\qquad\qquad
\omega_h^2 \simeq 4 \Omega^2 + k^2v_{th}^2,  
\end{equation}}
whereas in the opposite case of  very short wavelengths $(\rho_i^2 \sim d_i^2, \; k\rightarrow \infty$  where, however, a fluid-type description is not expected to hold  as mentioned at the end of this section)  we have 
{\begin{equation}\label{eq:L_fluid_limit2} 
\omega_l^2 \simeq \displaystyle{\frac{k^2v_{th}^2}{ 2}},
\qquad\qquad
\omega_h^2 \simeq 
 k^2\left(c_a^2+\displaystyle{\frac{3}{2}v_{th}^2} \right) . 
\end{equation}}
Note that in this formal limit the off-diagonal terms of the matrix tend to zero  and thus, as expected in the case of short wavelengths,  the two branches separate into a purely electrostatic $(\omega_h^2 \simeq 
 k^2 [c_a^2+ ({3}/{2})v_{th}^2])$  and a purely electromagnetic $(\omega_l^2 \simeq 
 k^2 v_{th}^2/2)$  mode.\\
For comparison with the CGL-FLR model discussed in Sec.\ref{CGL_FLR} it is interesting to  consider the first order corrections to the LFB, obtained in the small Larmor radius limit by  performing the ordering  $k^2\rho_i^2 \ll k^2d_i^2 $  and   $ k^2\rho_i^2\ll |4- k^2d_i^2|$. We find
\begin{equation}\label{eq:L_fluid_limit3} 
{\omega}_l^2 \simeq k^2\left[ c_a^2\left(1  
- \frac{k^2\rho_i^2}{2(4-k^2d_i^2)}
 \right)
+v_{th}^2 \left( 1 
 -k^2\rho_i^2
\frac{4-2k^2d_i^2}{\left( 4-k^2d_i^2\right)^3}  \right) \right] .
\end{equation} 
The inequality $ k^2\rho_i^2\ll |4- k^2d_i^2|$  follows from the  fact that a  small $k^2\rho_i^2$ expansion  is not valid in the vicinity of the  near crossing between the two branches mentioned above. 

We conclude this Section by noting that, when compared to the  corresponding dispersion relations computed from the truncated Vlasov expansion  in  Sec.\ref{Vlasov}, the range of validity of the dispersion relations represented in Fig.\ref{Fig_disp_tensor-0} will be restricted to values  $k\rho_i\leq 1$. Since  $k\rho_i=kd_i\sqrt{\beta}$, this condition becomes  more restrictive with increasing values  of $\beta$ when expressed in terms of $d_i$. Beyond the interval $kd_i\sqrt{\beta}\leq 1$  it is not possible to provide an  interpretation of the fluid hierarchy closure  in terms of an expansion of the Bessel functions (see Sec.\ref{Vlasov}-\ref{beta}).  Clearly the fluid description is expected to fail for $kd_i\gg 1$,  with its range of validity depending on the branch considered.

\subsection{Compressional effects and fluctuations of $B_z$}\label{B_z}
Both branches induce fluctuations in $\tilde{B}_z$ and $\tilde{\Pi}_{zz}$.  These  fluctuations are coupled through the relationship $\tilde{\Pi}_{zz}/\tilde{B}_z=P_{||}^0/B_0$, since linearisation gives 
\begin{equation}\label{eq:L_fluid_compress_1}
\frac{\tilde{\Pi}_{zz}}{P_{||}^0}=\frac{k\tilde{u}_x}{\omega},\qquad\qquad 
\frac{\tilde{B}_{z}}{B_0}=\frac{k\tilde{u}_x}{\omega}.
\end{equation}
This polarisation is coherent with an isothermal closure for the parallel ``temperature'', $\tilde{\Pi}_{zz}/(\tilde{n}_0m)$, where $\tilde{n}/n_0=k\tilde{u}_x/\omega $ is obtained from linearisation of Eq.(\ref{eq:M_1}), since for ${\bm k}=k_x{\bm e}_x$ the double adiabatic equation for $P_{||}$ (Eq.(\ref{eq:C_5})) corresponds to a polytropic with index 
$\Gamma_{||}=1$.

\subsection{Dispersion relation in the limit of a CGL closure with first order FLR}\label{CGL_FLR}
For further comparison we  consider the well known CGL fluid model with first order FLR corrections \cite{Thompson,Macmahon1,Macmahon2,Yajima,Chhajlani,Khanna} recently re-derived in  Ref.\cite{Cerri_2013} by starting from a full-pressure tensor model and by assuming $\omega/\Omega \sim kv_{th}/\Omega\sim\varepsilon\ll 1$. By construction this model projects the linear system of Eq.(\ref{eq:L_fluid_3}) onto the  LFB, since it relies on the small frequency approximation $\omega/\Omega\ll 1$ at the basis of the CGL closure. 
However, as first  noticed { in Ref.\cite{Macmahon3} and in Ref.\cite{Fredricks},} it fails to  describe  this magnetosonic branch  correctly for the purely perpendicular propagation that we are  considering here. This fact was further pointed out { in Ref.\cite{Mikhailovskii} } while discussing how the CGL model with first order FLR corrections may lead to  an  incorrect description of transversely propagating, low-frequency magnetosonic solitons. 
Here we remark  that the failure of the fluid description of fast magnetosonic  waves in the CGL-small FLR limit may be understood  by noticing that for ${\bm B}^0\cdot{\bm k}=0$ the CGL closure with first order FLR corrections does
 not take  account of the correct polarisation of the magnetosonic waves which requires  $\tilde{u}_y/\tilde{u}_x\sim kd_i\ll 1$.
For ${\bm B}^0\cdot{\bm k}=0$, indeed, the maximal ordering $\tilde{u}_x/ v_{th}\sim \tilde{u}_y /v_{th} \sim \varepsilon^0$, which the CGL-first order FLR model relies on, is inconsistent with the neglect of $\varepsilon^2$ corrections in Eq.(\ref{eq:linear_tens_xy}) because from Eq.(\ref{eq:linear_uy}) it follows that  $\tilde{u}_y / v_{th} \sim \varepsilon$. On the other hand,  the long wave-length limit, $k^2 d_i^2\sim \omega^2/\Omega^2 \ll 1$,  of Eq.(\ref{eq:L_fluid_limit3}), 
\begin{equation}\label{eq:FLR_limit_omega} 
 {\omega}_l^2 \simeq k^2\left[ c_a^2\left( 1- \frac{k^2\rho_i^2}{8}\right) 
+v_{th}^2 \left( 1 
 -\frac{k^2\rho_i^2}{16}  \right) \right],
\end{equation}
\begin{equation}\label{eq:FLR_limit_vg}
\frac{v_{l}}{c_a} = \frac{kc_a}{\omega_{l}}
\left\{\displaystyle{ 1-\frac{k^2\rho_i^2}{4}+\frac{\rho_i^2}{d_i^2}
\left(1-\frac{k^2\rho_i^2}{8}\right)}\right\},
\end{equation}
is directly obtained from the set of
 Eqs.(\ref{eq:linear_ux}-\ref{eq:linear_tens_yy}) after ordering $\omega/\Omega\sim kv_{th}/\Omega\sim\tilde{u}_y/v_{th}\sim\varepsilon$ and $\tilde{u}_x/v_{th}\sim\varepsilon^0$.  
  The relevant calculations are detailed in Appendix \ref{Appendix_C}, while in Sec.\ref{Comparison} there is the dispersion relation (Eq.(\ref{eq:Comparison_1})) to be compared  {with the small-$\beta$ limit of Eq.(\ref{eq:FLR_limit_omega}).   }
Here we recall that by linearising  the CGL-first order FLR set of equations for perpendicular propagation we would obtain instead 
\begin{equation}\label{eq:CGL_2}
\omega_{l}^2=
 k^2\left[c_a^2+v_{th}^2\left(1+\frac{k^2\rho_i^2}{16}\right)\right],
\end{equation}
with group velocity
\begin{equation}\label{eq:CGL_3}
\frac{v_{l}}{c_a} = \frac{kc_a}{\omega_{l}}
\left\{\displaystyle{ 1+\frac{\rho_i^2}{d_i^2}
\left(1+\frac{k^2\rho_i^2}{8}\right)}\right\}.
\end{equation}
  
\begin{figure}
\centerline{\epsfig{figure=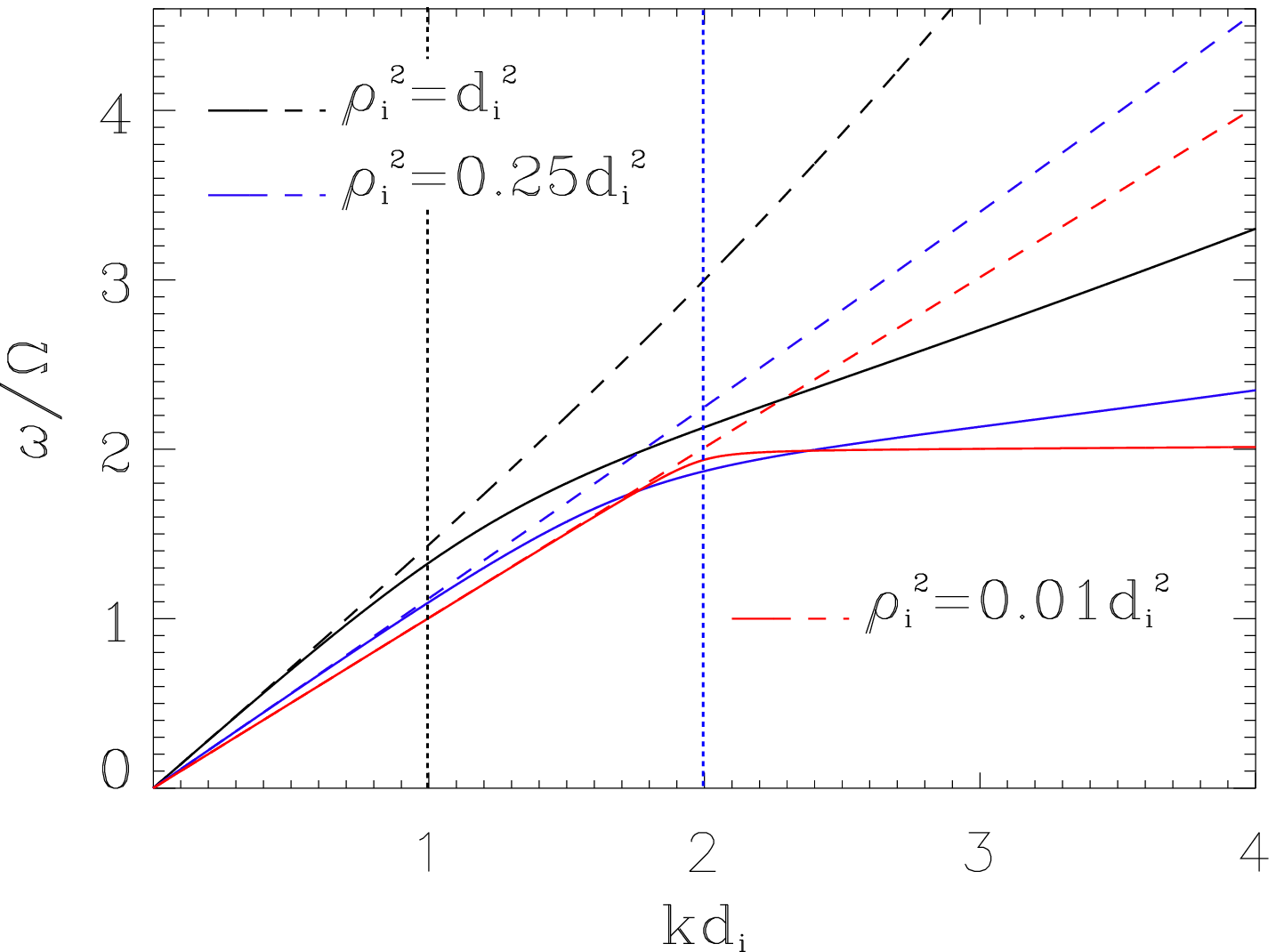,width=8.cm}
\epsfig{figure=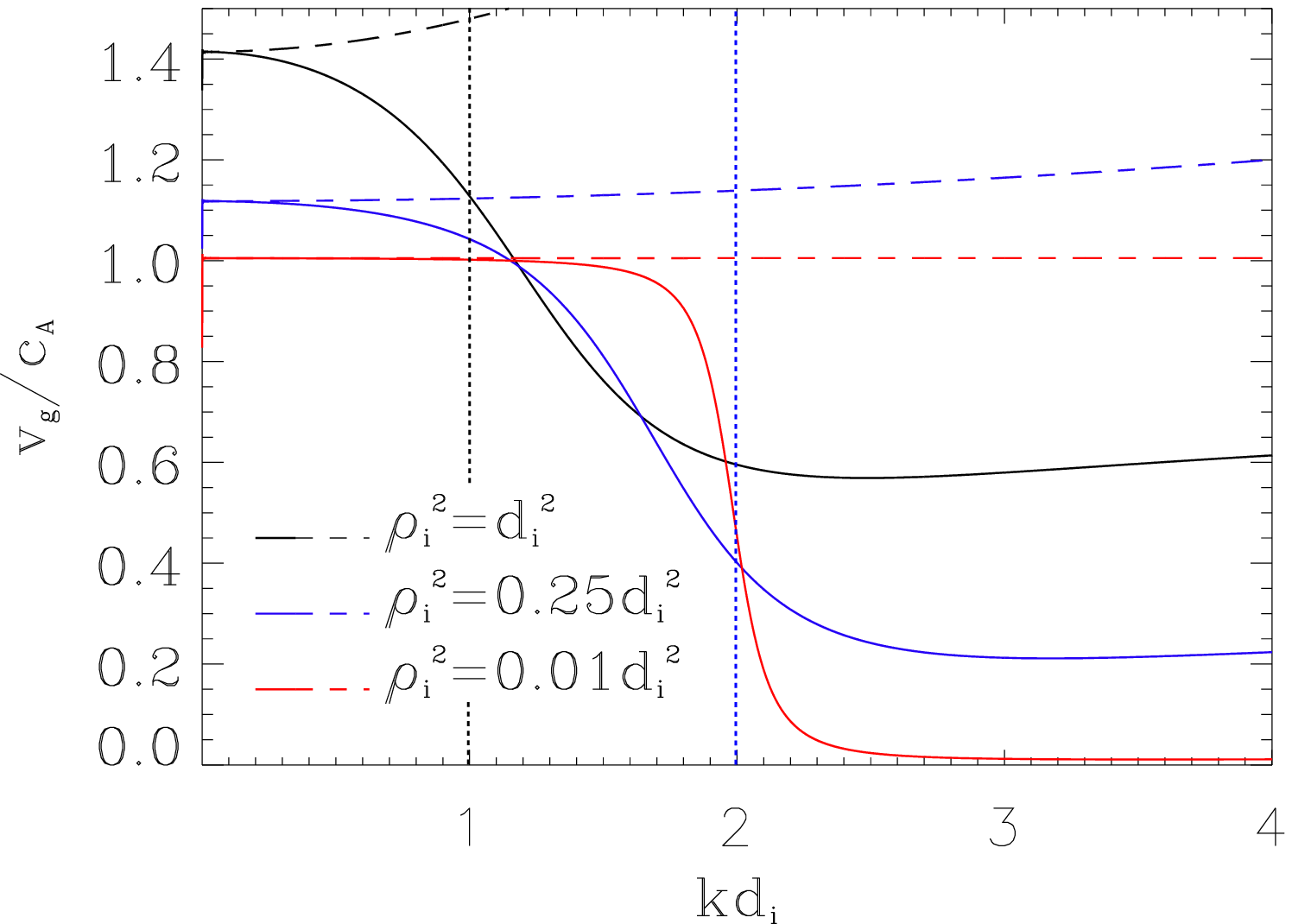,width=8.cm}}
\caption{Dispersion relations (left frame) and group velocities  versus $k d_i$ (right frame) for the  LFB as obtained  by retaining  small FLR corrections to the CGL closure for different values of $v_{th} / c_{a} = \rho_i  / d_i =\sqrt{\beta}$:  the frequency $\omega_l $ from Eq.(\ref{eq:CGL_2}) and  the group velocity $v_l$ from Eq.(\ref{eq:CGL_3}) are represented by dashed lines.  For comparison with the full pressure tensor model,  the curves of  $\omega_l $ and $v_l$  obtained from Eqs.(\ref{eq:L_fluid_5},\ref{eq:L_fluid_7}) are  also represented (solid lines) for the same parameters. For the red dashed curves ($\beta=10^{-2}$)  the condition $k\rho_i< 1$, required by the small-FLR expansion, is fulfilled over the whole interval. For the black ($\beta=1$) and blue ($\beta=1/4$) dashed curves the formal ranges of validity $kd_i< 1$ and $kd_i< 2$ are delimited by the  black and blue dotted vertical lines, respectively. Even if the above intervals also define the range in which curves with equal colors should be compared,  no formal restriction on the wavelength range is assumed for the solid lines  which are computed from the full pressure tensor fluid model without expanding for $k\rho_i< 1$ .}  \label{Fig_disp_tensor}
\end{figure}

This latter dispersion relation, first given in Ref. \cite{Yajima} {and obtained by neglecting the leading order \emph{perpendicular} (to ${\bm B}$)  heat-flux contribution as  for Eqs.(\ref{eq:FLR_limit_omega}-\ref{eq:FLR_limit_vg})}, misses 
 the FLR correction to the Alfv\'enic contribution of Eq.(\ref{eq:FLR_limit_omega}) because of the incorrect polarisation (Appendix \ref{Appendix_C}). In addition  the FLR correction to the thermal velocity appears  with the opposite sign. While
the $\sim k^2v_{th}^2k^2\rho_s^2$ term  should be neglected in the small-$\beta$ limit $k^2\rho_i^2\ll k^2d_i^2$, in which Eq.(\ref{eq:FLR_limit_omega}) coincides with the kinetic counterpart  obtained by consistently neglecting the heat-flux gradient contribution (Sec.\ref{Comparison}), in Eq.(\ref{eq:CGL_2})  it represents the only dispersive effect, which is of the same order  
as  the neglected perpendicular heat-flux term.\\
In Figs.(\ref{Fig_disp_tensor}) the dispersion relation for the LFB (left frame) and the dependence of the group velocity on  $k$ (right frame) at relatively small values of $k\rho_i$ are compared for the solution obtained in the fluid model with a full pressure tensor equation (Eqs.(\ref{eq:L_fluid_5},\ref{eq:L_fluid_7}), solid lines) and in  the CGL-FLR model (Eqs.(\ref{eq:CGL_2}-\ref{eq:CGL_3}), dashed lines) { for different values of $\rho_i/d_i$. The range of values of $kd_i$ in Figs.(\ref{Fig_disp_tensor}), corresponding  to values of $\rho_i$ that extend beyond the limit $k\rho_i\sim 1$, has been chosen in order to highlight the differences between the full pressure tensor model and the CGL-FLR limit.  This applies in particular to the cases $\rho_i=0.5 d_i$ (blue curves) and $\rho_i=d_i$ (black curves) in which the small FLR limit is often considered, especially for astrophysical applications (see e.g. the discussion in Ref.\cite{Cerri_2013}).}  Even if the differences  may appear negligible in the wave-length range $kd_i\ll 2$ and $k\rho_i\ll 1$,
a very different behaviour is  displayed for $kd_i\gtrsim 1$, even if $k^2\rho_i^2\rightarrow 0$ (red curves). These differences become remarkable for the group velocities, as the small FLR limit obtained from Eq.(\ref{eq:L_fluid_limit3})  gives a group velocity, see Eq.(\ref{eq:FLR_limit_vg}),  that decreases with increasing values of $kd_i$, a behaviour   in agreement  with the kinetic result \cite{Macmahon3,Fredricks,Mikhailovskii} (cf. Eq.(\ref{eq:Comparison_1}) next) and in contrast  with the increase  described by Eq.(\ref{eq:CGL_3}).

\subsection{Eigenmodes in the long wave-length limit}\label{eigenvectors}

In the long  wavelength limit $kd_i\ll 1$  the eigenvectors ${\bm V}^l( k,\omega_l)$ and ${\bm V}^h( k,\omega_h)$ in the perturbation space $\{
\tilde{u}_x, ~\tilde{u}_y, \\ \tilde{\Pi}_{xy}v_{th}/P^0_\perp, ~(\tilde{\Pi}_{xx}-\tilde{\Pi}_{yy})v_{th}/P^0_\perp
\}$  of the LFB  and of the HFB are respectively 
\begin{equation}\label{eq:vec_pol_LFB}
{\bm V}^l=~\tilde{u}_x^l   \left\{ 1, ~~i\frac{\Omega}{\omega_l}\frac{k^2\rho_i^2}{4},
 ~  \frac{\Omega}{kv_{th}}\frac{k^2\rho_i^2}{4},
~ - \frac{k^2\rho^2}{4\omega_l} \left(\frac{2\omega_l^2-k^2v_{th}^2}{kv_{th}}\right)
\right\}\qquad  \mbox{(LFB)},
\end{equation}
\begin{equation}\label{eq:vec_pol_HFB}
{\bm V}^h=~\tilde{u}_x^h   \left\{ ~1, ~-i, ~ - iv_{th}\frac{\omega_h}{k},
~ \frac{1}{\Omega}\left(\frac{2\omega_h^2-k^2v_{th}^2}{kv_{th}}\right)\right\}\qquad \mbox{(HFB)}.
\end{equation}
where the relative sign between the  pressure perturbations and the velocity perturbations depends on the propagation direction i.e., on the sign of the phase velocity. 
Note that, because of the presence of the two branches and {of their} different propagation properties,  an initial  incompressible perturbation  of the form   ${\bm V}=\{0,~\tilde{u}_0(x), ~0, ~0\}$  can  generate  compressible flows
since  the difference in the propagation velocities  removes the  cancellation  that makes $\tilde{u}_x(x)=0$ at the initial time. This feature is central to the mechanism of  shear-induced  agyrotropy   discussed  in Ref.\cite{Tensor_Anis} and  cannot occur in the CGL-FLR limit where only the $\omega_l$ mode is present.

\section{Kinetic description of the low and of the high frequency branches}\label{Vlasov}

While the low frequency magnetosonic branch has its counterpart in the MHD plasma description, this is not the case for the high frequency branch which cannot be described by the MHD equations. In order to characterise  the origin  and nature of this latter propagation branch more precisely,  in this section we
 consider for both branches the corresponding kinetic  dispersion relations as obtained from the solution of the linearised Vlasov-Maxwell equations. We assume  again  purely perpendicular propagation.\\ First we note that in order to account for the formation of a non-gyrotropic  pressure tensor  in the $x$-$y$ plane (i.e., in the plane perpendicular to the  uniform magnetic field ${\bm B}^0=B_0{\bm e}_z$), in the solution of the linearised Vlasov  equation  we need to include the $m= \pm 2$ terms in the expansion of the perturbed ion distribution function into harmonics of the gyration angle $\alpha$ in phase space. The combination of the contributions of these
 terms  to the kinetic dispersion relation  leads to the  $\omega^2- 4 \Omega^2$ resonances that we have encountered in the fluid treatment.  For a similar reason,  having neglected {in the fluid analysis} the  ion heat flux  $\partial_i Q_{ijk}$, we need not include the  $m= \pm 3$ (and higher order) terms {in this kinetic treatment}.

Thus we compute the ion contribution to the permittivity tensor   $K_{ij}$ by retaining the $n=0$, $n=\pm 1$ and  $n=\pm 2$   terms  in the  summation over the cyclotron harmonics  (see e.g. Ref.\cite{Krall}, p.404-405). { In addition,  in order to make the comparison with the fluid treatment meaningful, we  take the small Larmor-radius limit by retaining} contributions up to the power $\sim k^4d_i^2\rho_i^2$ while neglecting those in $\sim k^4\rho_i^4$. This means that in the elements of the permittivity  tensor  we have kept up to  the linear contributions in
$(k\rho_i)^2/2$ {and that the comparison with the fluid derivation is restricted to values $(k^2\rho_i^2)/(k^2d_i^2)=\beta\ll 1$}. Clearly this expansion procedure does not allow us to recover in the low frequency limit the  FLR correction to the $k^2v^2_{th}$ in Eq.(\ref{eq:FLR_limit_omega}). {On the other hand, it elucidates the inconsistency of the dispersive effects in the CGL-FLR dispersion relation Eq.(\ref{eq:CGL_2}).  These specific points} will be considered later in this Section.

We compute the electron  contribution to the permittivity tensor taking  the limits $\omega^2/\Omega_e^2 \to 0$  and $k^2\rho_e^2 \to 0$, with  {$\Omega_e= eB_0/(m_e c)$} and $\rho_e$ the electron cyclotron frequency and thermal Larmor radius respectively, as consistent with the zero mass (cold) limit assumed in the fluid treatment.  Thus  we  retain only the  off-diagonal contributions to  the permittivity tensor  of the ${\bm E}\times{\bm B}$-drift which are given  by the low frequency limit ($\omega^2/\Omega_e^2\ll 1$) of the $n=\pm 1$  terms in the  summation over the  electron cyclotron harmonics.
By adding the ion and the electron contributions  in the limits described above  we  obtain 

{ \begin{equation}\label{eq:model_V_1}
K_{11}=1 
-\,  
\underbrace{{\omega_{pi}^2}\,\frac{1-k^2\rho_i^2/2}{\omega^2-\Omega_i^2}}_{n=\pm 1} \,-\, \underbrace{{\omega_{pi}^2}\,
\frac{k^2\rho_i^2}{2(\omega^2-4\Omega_i^2)}}_{n=\pm 2},
\end{equation}

\begin{equation}\label{eq:model_V_2}
K_{12}= -K_{21}= 
{\underbrace{ - i \, \frac{\omega_{pe}^2}{\omega\Omega_e}}_{n=\pm 1\;(\alpha=e)}}  
\,-\,  \underbrace{i\,\frac{\omega_{pi}^2}{\omega}\Omega_i\,\frac{1-k^2\rho_i^2}{\omega^2-\Omega_i^2}}_{n=\pm 1} \,-\, \underbrace{i\,\frac{\omega_{pi}^2}{\omega}\Omega_i\,\frac{k^2\rho_i^2}{\omega^2-4\Omega_i^2}}_{n=\pm 2},
\end{equation}

\begin{equation}\label{eq:model_V_3}
K_{22}=1  \,
-\,\underbrace{\frac{k^2v_{th}^2}{\Omega^2}
\frac{\omega_{pi}^2}{\omega^2}}_{n=0} \,-\,
\underbrace{{\omega_{pi}^2}\,\frac{1-3k^2\rho_i^2/2}{\omega^2-\Omega_i^2}}_{n=\pm 1} \,-\,\underbrace{{\omega_{pi}^2}\,
\frac{k^2\rho_i^2}{2(\omega^2-4\Omega_i^2)}}_{n=\pm 2},\end{equation}

\begin{equation}\label{eq:model_V_4}
K_{33}=1  \,
-\,\underbrace{\left(1-\frac{k^2v_{th}^2}{2\Omega^2}\right)\frac{\omega_{pi}^2}{\omega^2}}_{n=0} \,-\,
\underbrace{{\omega_{pi}^2}\,\frac{k^2\rho_i^2}{2(\omega^2-\Omega_i^2)}}_{n=\pm 1}. 
\end{equation} }
where $\omega_{pe}^2/\Omega_e =  \omega_{pi}^2/\Omega_i$. The order $n$ of the cyclotron harmonic from which  each term comes  is labelled in the underbrace.
Note that the cancellation between the electron and the ion ${\bm E}\times{\bm B}$ contributions to the plasma current leads in the low frequency limit   to a partial cancellation between the  first two  terms in $K_{12},$ and $ K_{21}$.

Neglecting the displacement current, as consistent with the fluid model used above, i.e., by taking $c_a^2 \ll c^2$,  we write the normalised dielectric tensor 
$D_{ij} =  (\Omega^2/\omega_{pi}^2)  \{K_{ij}-(k^2\delta_{ij}-k_ik_j)c^2/\omega^2 \}  $ as

\begin{equation}\label{eq:L_Vlasov_1}
[{\bm D}]\equiv\left(
\begin{array}{ccc}
\displaystyle{\frac{\Omega^2-k^2 v_{th}^2/2}{\Omega^2-\omega^2}
+\frac{k^2v_{th}^2}{2(4\Omega^2-\omega^2)}}
 & \qquad 
\displaystyle{ i\frac{\Omega}{\omega}
\left(
\frac{\omega^2-k^2v_{th}^2}{\Omega^2-\omega^2}
+ \frac{k^2v_{th}^2}{4\Omega^2-\omega^2}
\right) } & \qquad 0
  \\ \\
-\displaystyle{ i\frac{\Omega}{\omega}
\left(
\frac{\omega^2-k^2v_{th}^2}{\Omega^2-\omega^2}
+ \frac{k^2v_{th}^2}{4\Omega^2-\omega^2}
\right) }
& 
 \quad \displaystyle{
-\frac{k^2(c_a^2+v_{th}^2)}{\omega^2}+
\frac{\Omega^2-3k^2v_{th}^2/2}{\Omega^2-\omega^2}
+\frac{k^2v_{th}^2}{2(4\Omega^2-\omega^2)}} & \qquad 0
\\ \\
0 & 0 &
{D}_{33}
\end{array}
\right).
\end{equation}
{ The vanishing of the coefficient 
\begin{equation}\label{eq:Vlasov_ordinary_1}
{D}_{33}\equiv\displaystyle{ 
\frac{k^2v_{th}^2}{2(\Omega^2-\omega^2)}
-\frac{\Omega^2+k^2(c_a^2-v_{th}^2/2)}{\omega^2}
}
\end{equation}
yields the dispersion relation of the so-called Ordinary mode, 
\begin{equation}\label{eq:Vlasov_ordinary_2}
\omega^2=\Omega^2-\frac{k^2v_{th}^2}{2(1+k^2d_i^2)},
\end{equation}
 which corresponds to electromagnetic perturbations with $\tilde{\bm E}$ parallel to ${\bm B}_0$. This mode has no counterpart in the fluid  description 
 in Sec. \ref{Linear_fluid} where, as already mentioned,  the ``hydrodynamic'' root  related to $\tilde{u}_z$  in Eq.(\ref{eq:L_fluid_3}) does not apply to  a collisionless 
plasma.

The dispersion relation of the transversely polarised modes, obtained from the vanishing of the determinant of Eq.(\ref{eq:L_Vlasov_1}) by consistently neglecting the $\sim k^4\rho_i^4$ contributions while keeping the $\sim k^4d_i^2\rho_i^2$ terms, is

\begin{equation}\label{eq:L_Vlasov_2}
 \frac{ \displaystyle{
\left(4-\frac{\omega^2}{\Omega^2}\right)
\left[\frac{\omega^2}{\Omega^2} -k^2d_i^2
\left(1-\frac{k^2\rho_i^2}{2}\right)\right]
-\frac{k^2\rho_i^2}{2}
\left[k^2d_i^2\left(1-\frac{\omega^2}{\Omega^2}\right)+8\right] } }
{\omega^2(\Omega^2-\omega^2)(4\Omega^2-\omega^2)}=0. 
\end{equation}
Its roots, whose behaviour  versus $k d_i$  is sketched in the left frame of Fig.\ref{Fig_Vlasov}, are (same notation as in Eq.(\ref{eq:L_fluid_5})) 
\begin{equation}\label{eq:L_Vlasov_3}
\left(\frac{\omega_{h,l}}{\Omega}\right)^2 = 2 + \frac{k^2d_i^2}{2}
 \pm\sqrt{ 4 - 2k^2d_i^2 -4k^2\rho_i^2 + 
\frac{k^4d_i^2}{4} \left(d_i^2+6\rho_i^2\right) }. 
\end{equation}
  The group velocities, represented in Fig.(\ref{Fig_Vlasov}), right frame, are given by
\begin{equation}\label{eq:L_Vlasov_4}
\frac{v_{h,l} }{c_a} = \frac{kc_a}{\omega_{h,l} }
\left\{\displaystyle{ \frac{1}{2}
\pm \frac{A }{B } }\right\} ,
\end{equation}
where 
\begin{equation}\label{eq:L_Vlasov_5} 
A  = -1-2\frac{\rho_i^2}{d_i^2}
+\frac{k^2d_i^2}{4}\left(1+6\frac{\rho_i^2}{d_i^2} \right), 
\end{equation}
\begin{equation}\label{eq:L_Vlasov_6} 
B  = \sqrt{ 4- 2k^2d_i^2 - 
4k^2\rho_i^2+\frac{k^2d_i^2}{4} \left(k^2d_i^2+6\,k^2\rho_i^2\right) }.
\end{equation}
As in the fluid description, see  Eqs.(\ref{eq:L_fluid_4}-\ref{eq:L_fluid_9}), in the limit $\rho_i=0$  the HFB vanishes as indicated by the  cancellation of the $4\Omega^2-\omega^2=0$  terms in the numerator and in the denominator of Eq.(\ref{eq:L_Vlasov_2}).  \\
The LFB and HFB described by the roots of Eq.(\ref{eq:L_Vlasov_3}) are respectively the  magneto-acoustic mode and a generalized   ion-Bernstein mode. In  the quasi-neutral limit considered here this mode is  not quasi-electrostatic, as consistent with the  general dispersion relation discussed in 
 Refs.\cite{Fredricks}, \cite{Fredricks2}.

\begin{figure}
\centerline{\epsfig{figure=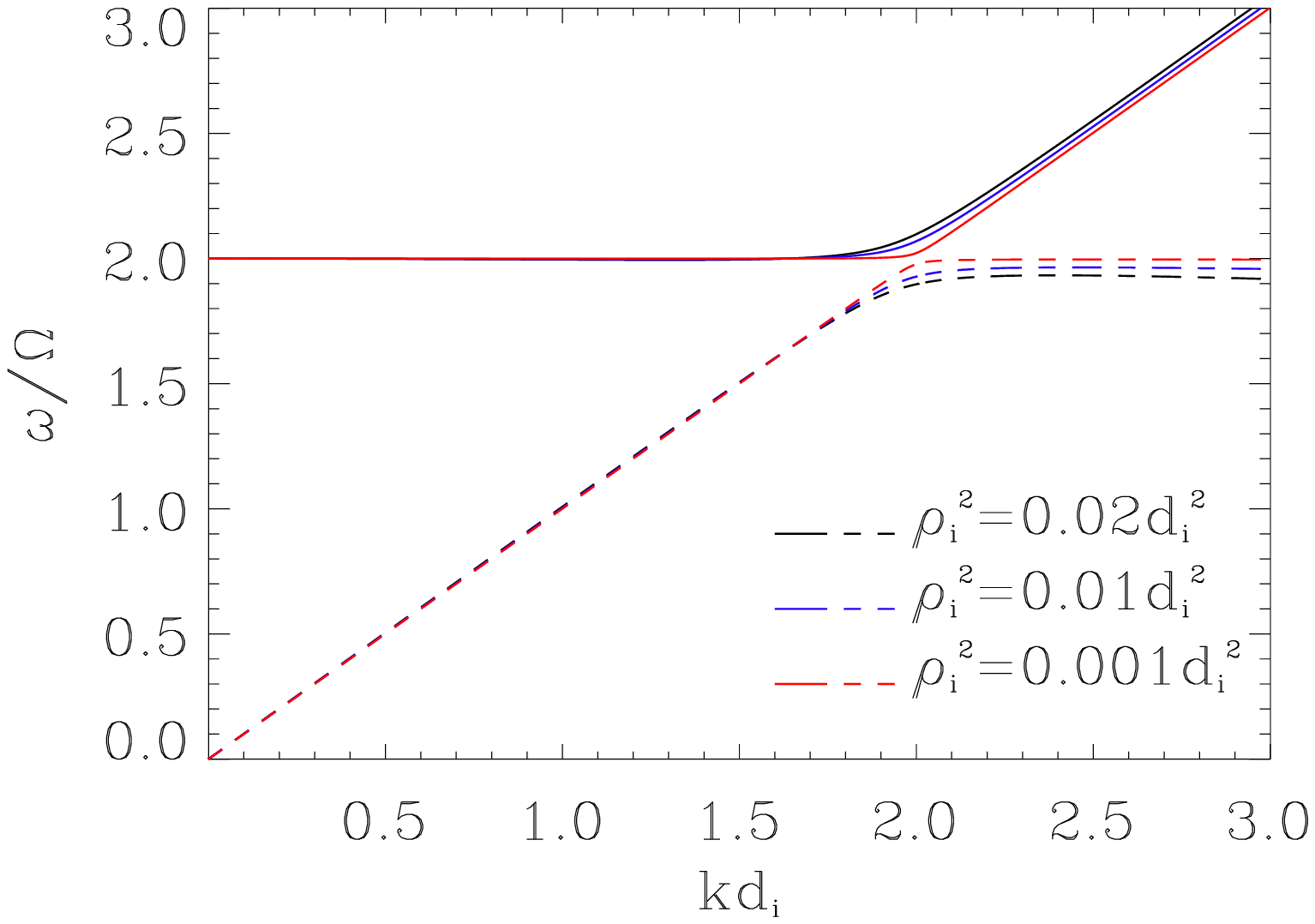,width=8.cm}
\epsfig{figure=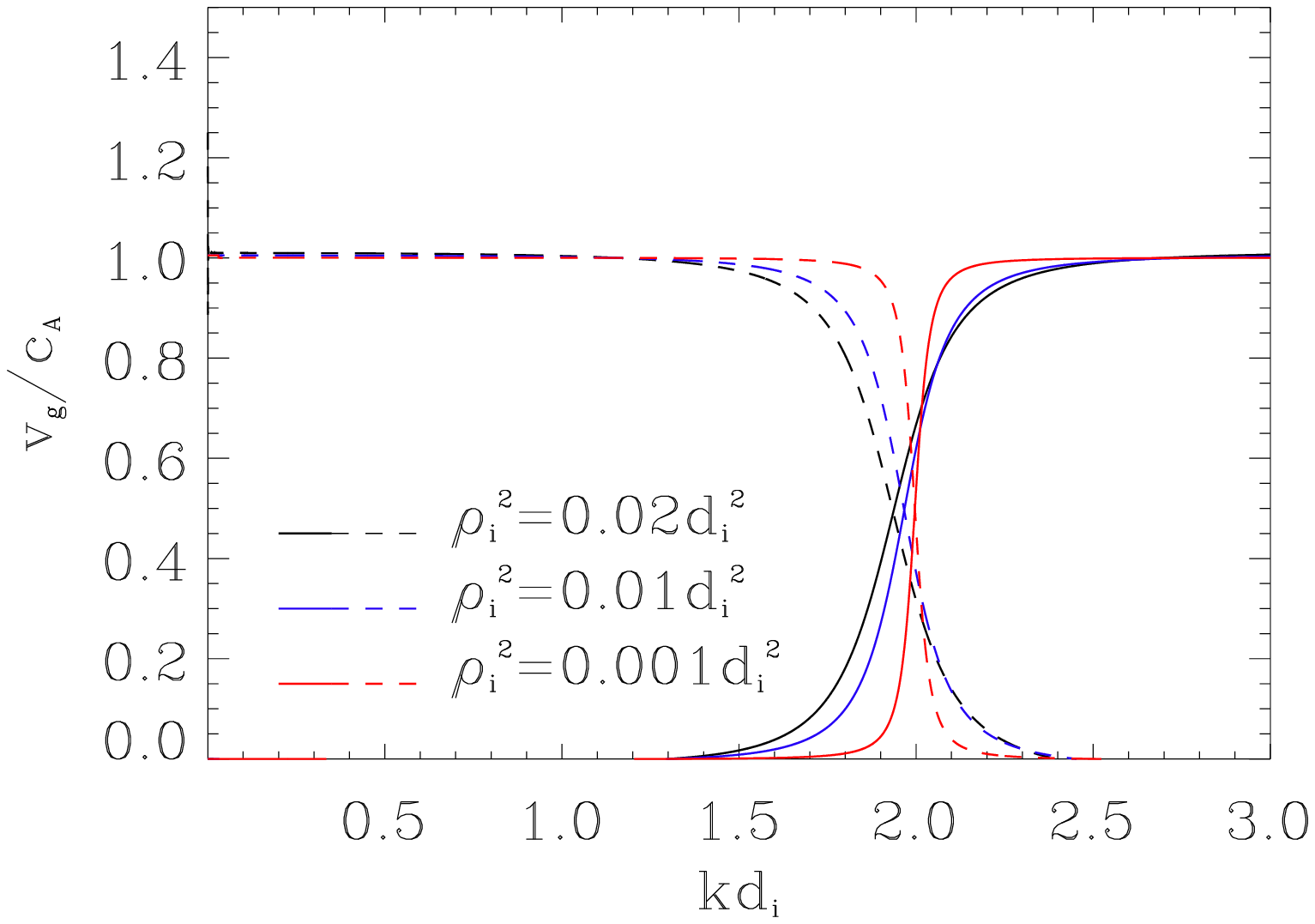,width=8.cm}}
\caption{  {Dispersion relations (left frame) and group velocities   versus $k d_i$   (right frame) of the  LFB ($\omega_l$ and $v_l$, dashed lines)  and of  the HFB ($\omega_h$ and $v_h$, solid lines) obtained from the  truncated Vlasov-Maxwell system for different values of $v_{th}/c_{a}=\rho_i/d_i{=\sqrt{\beta}} < 1$.  }}\label{Fig_Vlasov}
\end{figure}

\subsection{Comparison of the fluid and kinetic dispersion relations}\label{Comparison}
 Because of the assumptions made in the derivation of the kinetic dispersion relations (Eqs.(\ref{eq:L_Vlasov_2}-\ref{eq:L_Vlasov_6})), a meaningful  comparison with the fluid counterparts (Eqs.(\ref{eq:L_fluid_4}-\ref{eq:L_fluid_9}))  is possible only for  small values of $k^2\rho_i^2$ i.e., for  relatively  small values of the ratio $\rho_i/d_i{=\sqrt{\beta}}$, and for a restricted range of values of $kd_i$.   This comparison is shown   in Fig.(\ref{Fig_Comparison}).\\

\begin{figure}
\centerline{\epsfig{figure=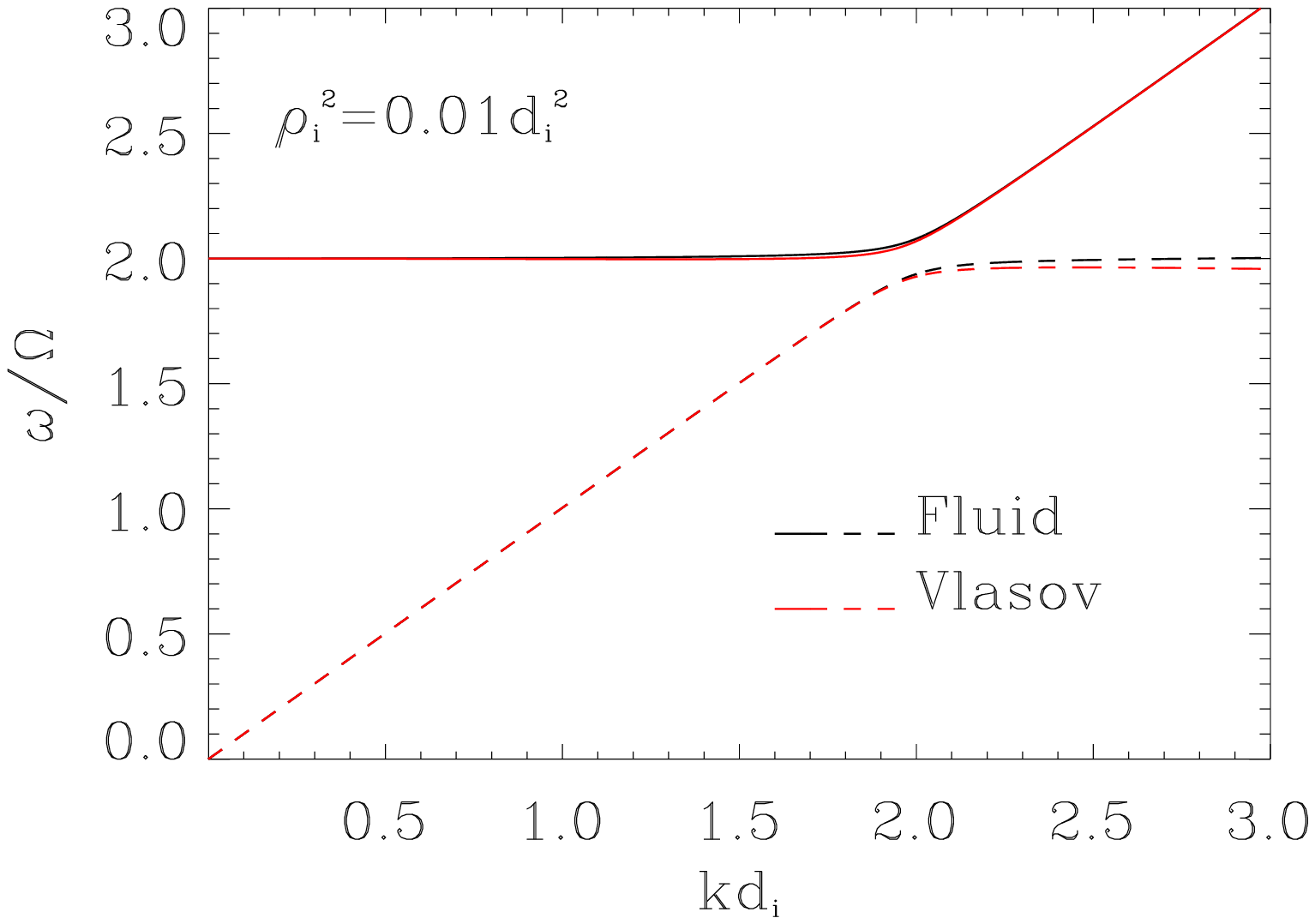,width=8.cm}
\epsfig{figure=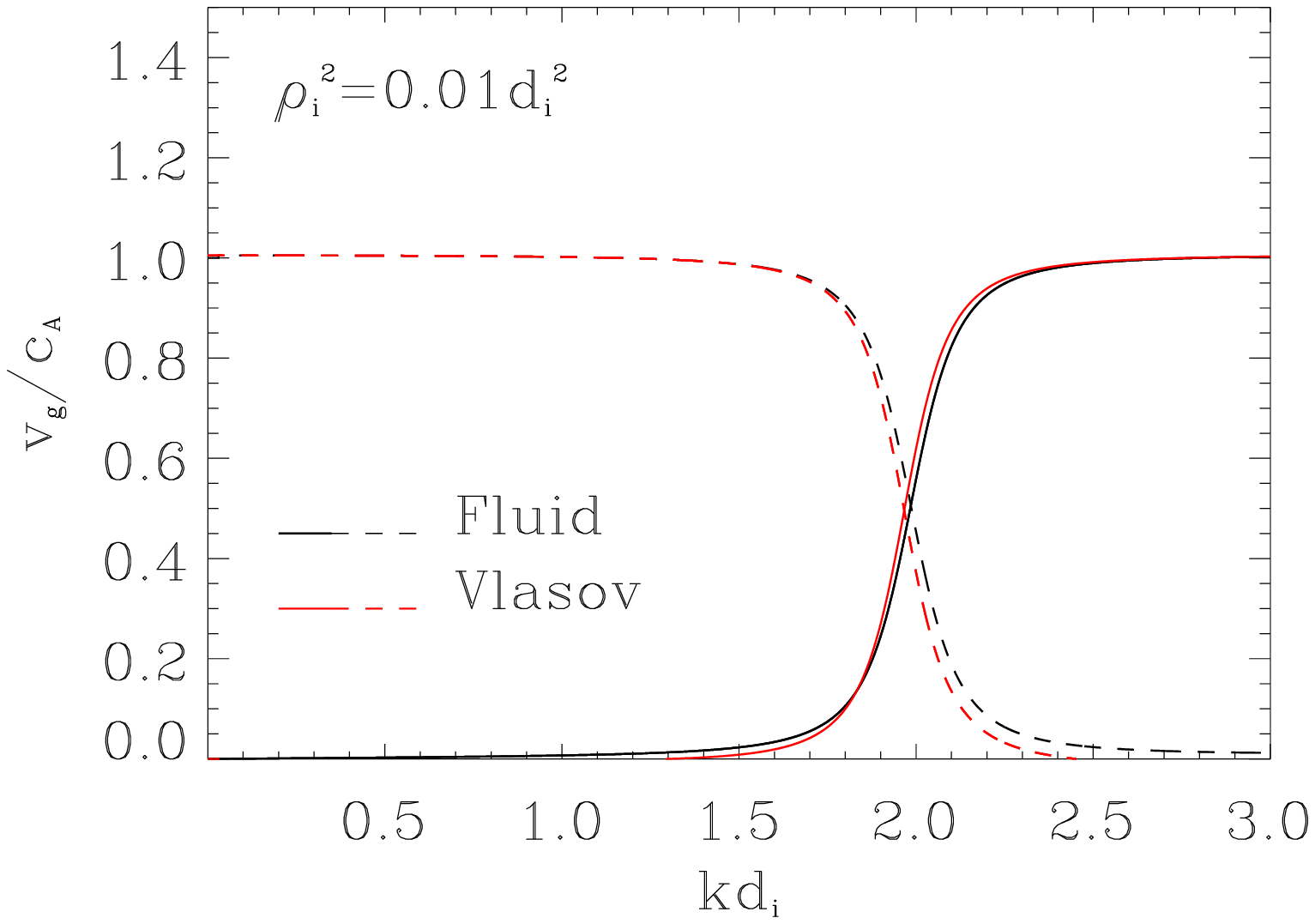,width=8.cm}}
\caption{  {Dispersion relations (left frame) and group velocities   versus $k d_i$   (right frame) of the  HFB $\omega_h$  obtained from the truncated Vlasov-Maxwell system (red lines)  and from the fluid model (black lines). Both the HFB (solid lines) and the LFB (dashed lines) are represented.  The value  $v_{th}/c_{a}=\rho_i/d_i{=\sqrt{\beta}} =0.01$ has been chosen as representative for the $k^2\rho_i^2\ll1$ limit in the  interval of wave-numbers considered.  }}\label{Fig_Comparison}
\end{figure}

First we note that in this small $k^2\rho_i^2$ limit  the HFB described by Eq.(\ref{eq:L_Vlasov_3})  has, near  the resonance $\omega=2 \Omega$, the  factorized expression  
\begin{equation}\label{eq:parabola_Vlasov}
 \frac{\delta \omega}{\Omega} \sim -  \frac{\rho_i^2}{d_i^2}\,\, f(k^2 d^2_i)=  - \frac {\rho_i^2}{d_i^2} \,\, \frac
{(  k^2 d_i^2  )\, ( {3}k^2 d_i^2/2 - 4 )}{2(4 -  k^2 d_i^2)}.
\end{equation}
We see that the dependence of $\delta \omega \equiv  \omega  - 2\Omega  $ on  $\rho_i^2$  appears only through  a multiplicative factor  whereas the values of $kd_i$  corresponding to  its minimum   (zero of the group velocity) and  to the  second crossing of the resonance at     $kd_i=2\sqrt{2/3}$, are independent of $\rho_i/d_i$.
On the contrary  in the fluid model the HFB remains  above the $\omega = 2\Omega$ resonance for all $k\not=0$.
This different behaviour is  made evident by  the  opposite spatial dispersion of the HFB for $k^2 \to 0$  in the two models   where we find 
\begin{equation}\label{eq:comp_2}
\omega_h^2\simeq 4\Omega^2\pm k^2v_{th}^2,  \qquad\qquad {v_{h} } \simeq
\pm \frac{kv_{th}^2}{2\Omega}, 
\end{equation}
where the plus sign refers to the fluid model and the minus sign to the kinetic treatment.
This  opposite  behaviour  is also shown by  the first order correction to  the ratio of the  perturbed pressure tensor components,
\begin{equation}\label{eq:comp_3}
{\tilde\Pi}_{xx} / {\tilde\Pi}_{yy} \simeq -(1 \pm k^2\rho_i^2),
\end{equation}
 where again the plus sign refers to the fluid model.  The   perturbed pressure  components   are computed  in the fluid model  directly from the linearised  pressure  tensor equations  and in the kinetic treatment from   the second order velocity moment of the perturbed distribution function (Ref.\cite{Krall}, Eq.8.10.8). 

Secondly we note that while in the fluid  model the LFB magnetosonic branch crosses the $\omega = 2 \Omega$ resonance at a value which in the limit $k^4\rho^4\ll 1$ is  $kd_i\simeq 2\sqrt{2}$ (as obtained from Eq.(\ref{eq:L_fluid_5})), in the kinetic treatment this branch stays below this resonance.
More importantly, having retained only the $k^2\rho_i^2$ corrections in both the diagonal and off-diagonal matrix elements of the dielectric tensor,   the  low frequency limit of the kinetic dispersion relation  (Eq.(\ref{eq:L_Vlasov_3})) can  account  consistently only for the  FLR corrections to $c_a^2$ in the dispersion relation of the  magnetosonic branch.

{The magnetosonic dispersion relation obtained by disregarding the $\sim k^4\rho_i^4$ terms in the expansion of  the LFB of Eq.(\ref{eq:L_Vlasov_3})  is } 
\begin{equation}\label{eq:Comparison_1} 
 {\omega}_l^2 \simeq k^2\left[ c_a^2\left( 1- \frac{k^2\rho_i^2}{8}\right) 
+v_{th}^2  \right].
\end{equation}
{This equation agrees  with the small-$\beta$ limit of the  kinetic result obtained in Refs. \cite{Fredricks,Mikhailovskii} and coincides with the consistent limit of Eq.(\ref{eq:FLR_limit_omega}).}

\section{Role of the heat-flux in the long  wave-length limit of the LFB for $\beta\sim 1$}\label{beta}

 Neglecting $\sim k^4\rho_i^4$ contributions in the power expansion of Eq.(\ref{eq:L_Vlasov_3}) allows us to obtain the same LFB dispersion relation  (\ref{eq:Comparison_1}) from  the truncated Vlasov-Maxwell system and from the fluid model, closed so as  to include the full pressure tensor equation without the divergence of the heat flux. This however does not imply that the heat-flux can always be  consistently neglected even when only terms that scale with  the  perpendicular wave number as $k^4$ are  included.

For $\beta$ of order unity  and for  purely perpendicular propagation, terms of order  $\sim k^2v_{th}^2k^2\rho_i^2$ must be retained  in the dispersion relation of the LFB  in  the long  wave-length limit.
In the fluid description  a term of this order  arises  from the heat flux  in the limit  $\omega\ll \Omega$ and needs to be retained  for a consistent description of the LFB, see Appendix \ref{Appendix_C}. This  feature of  the LFB was first pointed out and extensively discussed in Ref.\cite{Mikhailovskii}. In the kinetic derivation a term of the same order arises  from  the $n=0$ contribution to the $K_{22}$ element of the permittivity tensor which,  for  $\omega/\Omega \ll 1 $, requires  us  to retain an additional $\sim k^2\rho_i^2$ term in  the expansion of the  Bessel  function  since this term  contributes to the dispersion relation of the LFB  to  the same order of accuracy, in powers of $(\omega/\Omega)^2\sim k^2\rho_i^2$,  as  the other contributions to Eq.(\ref{eq:model_V_3}). In doing so Eq.(\ref{eq:model_V_3}) becomes
\begin{equation}\label{eq:K_22_corrected}
K_{22}=1  
-\,\underbrace{\frac{k^2v_{th}^2}{\Omega^2}\left(1-{\frac{3}{4}}k^2\rho_i^2\right)
\frac{\omega_{pi}^2}{\omega^2}}_{n=0} \,-\,
\underbrace{{\omega_{pi}^2}\,\frac{1-3k^2\rho_i^2/2}{\omega^2-\Omega_i^2}}_{n=\pm 1} \,-\,\underbrace{{\omega_{pi}^2}\,
\frac{k^2\rho_i^2}{2(\omega^2-4\Omega_i^2)}}_{n=\pm 2}\end{equation}
and the $D_{22}$ element of matrix (\ref{eq:L_Vlasov_1}) is rewritten as 
\begin{equation}\label{eq:D_22_corrected}
D_{22}=  \displaystyle{
-\frac{k^2\left[ c_a^2+v_{th}^2\left(1-\displaystyle{\frac{3}{4}k^2\rho_i^2} \right) \right]}{\omega^2}+
\frac{\Omega^2-3k^2v_{th}^2/2}{\Omega^2-\omega^2}
+\frac{k^2v_{th}^2}{2(4\Omega^2-\omega^2)}} 
.\end{equation}
This leads to the correct extension of Eq.(\ref{eq:Comparison_1}) including  $\sim k^2v_{th}^2k^2\rho_i^4$ terms, which reads \cite{Fredricks,Mikhailovskii}
\begin{equation}\label{eq:Comparison_2} 
 {\omega}_l^2 \simeq k^2\left[ c_a^2\left( 1- \frac{k^2\rho_i^2}{8}\right) 
+v_{th}^2 \left( 1 
 -\frac{5}{16}k^2\rho_i^2  \right) \right].
\end{equation}

Finally we note   that the $|n|=1$ harmonics are absent  from  the dispersion relation of   both the fluid and  the ``truncated'' Vlasov analysis.     In order to recover the $|n|= 1$  harmonics in the dispersion relation obtained from the truncated Vlasov derivation  terms proportional to $(k_\perp \rho_i)^4$ must be retained  and the limit to  $\omega \to \Omega$ and $(k_\perp \rho_i)^2 \to 0$ must be performed in an appropriate  order.  Correspondingly, it can be shown that the 
$|n|= 1$  harmonics are recovered within the fluid description if the divergence of the heat flux tensor is retained in the equation for the time evolution of the pressure tensor.  Consistently with the truncated Vlasov formulation this contribution is proportional to $(k_\perp \rho_i)^4$.

 \section{Conclusion}\label{Conclusion}

The aim of this article is to  characterise the linear modes that can propagate in a homogeneous plasma in the direction perpendicular to an externally imposed magnetic field within a two-fluid description  that uses a simplified model for the  cold electron fluid and a full pressure tensor treatment of the ion fluid. The perturbed velocity fields of  these modes   lie in the plane perpendicular to the magnetic field. This model description in its nonlinear version has been used  used in the investigation of the generation of a non gyrotropic pressure tensor \cite{Tensor_Anis}. 
 It is thus  important to have an {\it a priori} understanding of the physical  properties of this  model and of its limitations, in particular since it has been examined in full detail only in the context of its equilibrium configurations \cite{Cerri_2014}.
A first step in this line is to determine
 which kind of waves this model describes  and in particular how well they correspond to the results  obtained by solving the Vlasov equation. 
 Such a comparison is more easily performed   under the homogeneous plasma conditions adopted in the previous sections. \\
More specifically we have compared the linear dispersion relations  derived from this fluid model with those obtained from the solution of the Vlasov equation in a magnetised plasma by retaining, in the appropriate expansion in the thermal ion Larmor radius, the contribution of the $n=\,  0,\,\pm1,\, \pm2$ ion cyclotron harmonics. 
 The latter harmonics are required in order to account for the dynamics of an anisotropic, and in general non gyrotropic,  pressure tensor.  This comparison shows 
 that  plasma kinetic dynamics   involving the second cyclotron resonance can be mimicked, in the appropriate small Larmor radius limit, by a fluid description provided a full tensor pressure dynamics is retained.  Even if  the agreement between the two descriptions is not exact  in some cases, as explicitly mentioned in our analysis, the overall framework is coherent enough to make the extended fluid model viable and useful when  spatially   extended  inhomogeneous configurations are considered.  In such a case   fully kinetic simulations are exceedingly expensive in computational terms.  \\
Besides, this fluid approach makes it possible to identify,   more easily than with a full kinetic description, some features of the normal modes propagating in the system, such as the way the dynamical plasma response to perturbations determines  the  polarisation of these normal modes.

The inclusion of ion   thermal
  fluxes  in the plane perpendicular to the magnetic field  would extend this description   making it possible to recover  the  $n=\pm 1$ cyclotron harmonics that  do not appear in the dispersion relations derived  above and to include in principle higher harmonics  together with higher order terms in Bessel function expansion.  In addition, in Sec. \ref{beta} we have shown  that,  while  corrections arising from  higher order terms in Bessel expansion of the $n=0$ cyclotron harmonic appear to be  negligible  at high frequencies $\omega\gtrsim \Omega$ for small $k$, they  become important when considering FLR-corrections to the magnetosonic dispersion relation at non-negligible values of $\beta$  as evidenced in \cite{Fredricks, Mikhailovskii} in terms of the contribution of the perpendicular heat flux.   In this context of  comparison between fluid and kinetic frameworks, we recall  that the  dispersion relations of magnetised plasma modes  such as  kinetic magnetosonic modes and, using a different approach, Bernstein waves,   have been reconstructed  from a fluid-like analysis in Refs. \cite{Lau1,Lau2}.

More specifically we have shown  that the fluid model describes  two wave branches  that involve  velocity perturbations perpendicular to the magnetic field: a low frequency (LFB) and a high frequency (HFB) branch  with dispersion relations $\omega_l^2\simeq k^2 (c_{a}^2+v_{th}^2)$ and $\omega_h^2\simeq 4\Omega^2+k^2v_{th}^2$ respectively in the long wavelength limit $kd_i\rightarrow 0$.  While the LFB corresponds to the fast magnetosonic waves propagating in a  cold-electron plasma, where the thermal ion speed replaces the usual sound-velocity, the HFB finds no fluid counterpart in the isotropic pressure limit. The origin of the HFB can be understood by comparison with the kinetic  dispersion relation in the small FLR limit (Sec.\ref{Vlasov}), from which we recognise that the HFB  corresponds to   a generalization of the $2 \Omega$  ion-Bernstein mode.

Finally we underline that the  so called CGL-FLR fluid limit, obtained by expanding  to the lowest order FLR corrections the pressure tensor 
around a zeroth order  double adiabatic pressure {with no heat fluxes}, 
  misses the $k^2v_{th}^2$  corrections and  obtains  a coefficient different both in magnitude  and sign for the  $k^2v_{th}^2k^2\rho_i^2$  correction  as already evidenced in  Refs. \cite{Fredricks, Mikhailovskii} (see also  Appendix \ref{Appendix_C}). The source of this discrepancy is in the incorrect retaining of the magnetosonic branch polarisation in the CGL-FLR model   and in the subsequent neglect of the heat flux term, whose contribution needs to be retained  at the same order of accuracy.

\appendix

\section{Kinetic derivation of the two-fluid model with full pressure tensor dynamics}\label{Appendix_A}

Let us multiply  the Vlasov equation for the particle  distribution function $f ^\alpha({\bf x},{\bf v},t)$, with   $\alpha=e,i$ denoting  electrons and ions,
by a function $\chi({\bf x},{\bf v})$. Integration over $d^3v$, under 
 the  proper convergence conditions as  $|{\bf v}|\rightarrow\infty$, 
 leads  to the so-called ``phase-space conservation theorem'',
\begin{equation}\label{eq:A_1} 
\frac{\partial}{\partial t}  n ^\alpha \langle
\chi\rangle ^\alpha\,+\,\nabla_{x}\cdot  n ^\alpha\langle{\bf v}\,
\chi\rangle ^\alpha\,-\,n ^\alpha\,\langle{\bf v}\cdot\nabla_{x}\chi\rangle ^\alpha 
\,-\,
\frac{n ^\alpha}{m ^\alpha}\langle{\bf F} ^\alpha\cdot\nabla_{v} \chi \rangle ^\alpha\,=\,0,
 \end{equation}
where ${\bf F} ^\alpha({\bf x},t) =
 q ^\alpha({\bf E}
 ({\bf x},t)
 +{{\bf v}/c}\times{\bf B}
 ({\bf x},t))$
and we have introduced the  particle density $n ^\alpha({\bf x},t)=\int f ^\alpha({\bf x},{\bf v},t)d^3v$ and the mean value over the $\alpha$-species  distribution $\langle A({\bf x},t)\rangle ^\alpha\,=\,(1/n ^\alpha({\bf x},t))\,
 \int\,A({\bf x},{\bf
v},t)\,f ^\alpha({\bf x},{\bf v},t)\,d^{3}v$.  By introducing  the mean velocity ${\bf u} ^\alpha({\bf x},t)\equiv\langle {\bf v}\rangle ^\alpha$,  the continuity equation (Eq.(\ref{eq:A_2})) and  the Euler equation (Eq.(\ref{eq:A_3}))  are obtained from the zeroth ($\chi=m ^\alpha$) and  from the first order moment 
($\chi=m ^\alpha v_i$)  respectively. The  evolution  (Eq.(\ref{eq:A_4})) 
 of the pressure tensor $\Pi_{ij}^{\alpha}
({\bf x},t) 
=\,n ^\alpha m ^\alpha\langle v_{i}v_j\rangle ^\alpha-n ^\alpha m ^\alpha u_{i}^{\alpha}
u_j^{\alpha}$ is obtained from the the second anisotropic moment,
$\chi_{ij}=m ^\alpha v_iv_j$.
This equation involves the gradients of the  heat flux tensor
$Q_{kij}^{\alpha}\equiv \langle m ^\alpha n ^\alpha (v_k-u_k^{\alpha})(v_i-u_i^{\alpha})(v_j-u_j^{\alpha}) \rangle$.  No explicit  dependence on the electric field is present since the two  terms  $n^\alpha / m^\alpha (E_i u_j^\alpha + E_j u_i^\alpha)$ can be re-expressed in terms of the other variables by using  the momentum equation.  
Adopting a Cartesian tensor notation with lower indices for the spatial components, the three moment equations for the species $\alpha$ are:
\begin{equation}\label{eq:A_2}
\frac{\partial n^\alpha}{\partial t}\,
\,+\,\frac{\partial}{\partial x_i}( n^\alpha u_i^\alpha)\,=\,0,
\end{equation}
\begin{equation}\label{eq:A_3}
\frac{\partial u_i^\alpha}{\partial t}\,+\,u_k^\alpha \frac{\partial u_i^\alpha}{\partial x_k}\,=
\frac{q^\alpha}{m^\alpha c}(cE_i+\varepsilon_{ilm}u_l^\alpha B_m)\,
-\,\frac{1}{m^\alpha n^\alpha }\frac{\partial \Pi_{ik}^\alpha}{\partial x_k},
\end{equation}
\begin{equation}\label{eq:A_4}
\frac{\partial \Pi_{ij}^\alpha}{\partial t} \,+\,\frac{\partial Q_{kij}^\alpha}{\partial
x_{k}} \,+\,\frac{\partial }{\partial x_{k}}(u_k^\alpha\,\Pi
_{ij}^\alpha)\,+\, \frac{\partial u_{i}^\alpha}{\partial x_{k}}\Pi
_{kj}^\alpha\,+\,\frac{\partial u_{j}^\alpha}{\partial
x_{k}}\Pi_{ik}^\alpha \end{equation}
$$\,-\,\frac{q^\alpha}{m^\alpha c}
 \left(\varepsilon_{ilm}\Pi_{jl}^\alpha B_{m}\,+\,\varepsilon_{jlm}\Pi_{il}^\alpha B_{m} \right)\,=\,0.$$ 
In the latter $\varepsilon_{jlm}$ is the  Levi-Civita symbol in three dimensions. Provided some closure condition for  $Q_{ijk}^\alpha$ is given, the system of fluid equations above is closed once it is coupled to  the equations for the e.m. fields, 
\begin{equation}\label{eq:A_5}
\frac{\partial E_i}{\partial x_i}=\frac{1}{4\pi}(n^eq^e+n^iq^i),\qquad
\frac{\partial B_i}{\partial x_i}=0 ,
\end{equation}
\begin{equation}\label{eq:A_6}
\frac{\partial B_i}{\partial t}=-c\varepsilon_{ijk}
\frac{\partial E_k}{\partial x_j},
\end{equation}
\begin{equation}\label{eq:A_7}
\varepsilon_{ijk}\frac{\partial B_k}
{\partial x_j}=\frac{4\pi}{c}{J_i}+\frac{1}{c}\frac{\partial E_i}{\partial t},
\qquad
{J}_i\equiv n^eq^e{u}_i^e+n^iq^i{u}_i^i.
\end{equation}
Notice that the set of Eqs.(\ref{eq:A_2}-\ref{eq:A_7}) leads  to an energy conservation equation of the form
{\begin{equation}\label{eq:A_8}
\frac{\partial}{\partial t}\left\{\sum_\alpha\left[
\frac{n^{\alpha}m^{\alpha}}{2}(u^{\alpha})^{2} 
+\frac{\mbox{tr}\{{\bf \Pi}^\alpha\}}{2}\right]
+\frac{B^{2}}{8\pi}+\frac{E^{2}}{8\pi}\right\}=
\end{equation}
$$=-\nabla\cdot\left\{\sum_\alpha\left[\mathcal{\bm Q}^\alpha  +{\bf u}^\alpha\cdot{\bf \Pi}^\alpha+{\bf u}^\alpha\left(
\frac{\mbox{tr}\{{\bf \Pi}^\alpha\}}{2}+ \frac{n^\alpha m^\alpha
(u^{\alpha})^{2}}{2}\right) \right]
+\frac{c}{4\pi}{\bf
E}\times{\bf B}\right\} ,$$}
where on  the r.h.s. the heat flow vector $\mathcal{Q}_i^\alpha\equiv Q_{ijk}^\alpha\delta_{jk}/2$ has been introduced.

\section{Two-fluid double adiabatic CGL closure from the full pressure tensor equation}\label{Appendix_B}
 The CGL closure for the species $\alpha$ is obtained  from Eq.(\ref{eq:A_4}), in the limit of    a sufficiently strong magnetic field and/or sufficiently weak velocity strain, 
 by performing an expansion in powers of $1/( \tau_H^\alpha\Omega_\alpha)\ll 1$, where $\tau_H^\alpha\equiv |{\bm\nabla}{\bm u}^\alpha|^{-1}$ and $\Omega_\alpha\equiv |q^\alpha|B_0/(m^\alpha c)$,  and by  considering time scales long with respect to $\Omega_\alpha^{-1}$ such that 
${\partial}/{\partial t}  \sim 1/\tau_H^\alpha$. To leading order we obtain
\begin{equation}\label{eq:C_1}
\varepsilon_{ilm}\Pi_{jl}^{0,\alpha} B_{m}\,+\,\varepsilon_{jlm}\Pi_{il}^{0,\alpha} B_{m} 
\,=\,0,
\end{equation}
which gives 
\begin{equation}\label{eq:C_2}
\Pi_{ij}^{0,\alpha} =P_{\perp}^\alpha\delta_{ij}+(P_{||}^\alpha-P_{\perp}^\alpha)b_ib_j,
\end{equation}
where $P_{||}^\alpha$ and  $ P_{\perp}^\alpha$ are the parallel and perpendicular pressures, respectively. By contracting  {Eq.(\ref{eq:A_4})} 
with $\delta_{ij}$  we obtain \begin{equation}\label{eq:C_3}
\frac{d}{d t ^\alpha}(\Pi_{ij}^\alpha\delta_{ij})+
(\Pi_{ij}^\alpha\delta_{ij})\frac{\partial u_k^\alpha}{\partial x_k}
+2\Pi_{ik}^\alpha\frac{\partial u_i^\alpha}{\partial x_k}=0,
\end{equation} 
where 
\begin{equation}\label{eq:C_3b}
{\frac{d}{d t ^\alpha}\equiv\partial_t + u_k^\alpha\partial_k}.
 \end{equation}

Similarly,  contracting Eq.(\ref{eq:A_4})   with $b_ib_j$ gives 
 \begin{equation}\label{eq:C_4}
\frac{\partial }{\partial t}(\Pi_{ij}^\alpha b_ib_j)+
\frac{\partial }{\partial x_k}\left(u_k^\alpha
\Pi_{ij}^\alpha b_ib_j \right)
-\Pi_{ij}^{\alpha}\frac{\partial}{\partial t}
\left( b_ib_j\right)\end{equation} 
$$ +b_ib_j\left(\Pi_{ik}^\alpha\frac{\partial u_j^\alpha}{\partial x_k}
+\Pi_{kj}^\alpha\frac{\partial u_i^\alpha}{\partial x_k}
\right)=0.$$ 
Eqs.(\ref{eq:C_3}-\ref{eq:C_4})  are valid regardless of the relative ordering between  and $\tau_H^{\alpha}$ and 
$\Omega_{\alpha}^{-1}$ since  
$(\varepsilon_{ilm}\Pi_{jl}^{0,\alpha} B_{m}\,+\,\varepsilon_{jlm}\Pi_{il}^{0,\alpha} B_{m})$  disappears after  contraction with either  $\delta_{ij}$  or  $b_ib_j$. 
In the  double-adiabatic ordering, substituting  the leading order term $\Pi_{ij}^{0,\alpha}$  for $\Pi_{ij}^{\alpha}$  into Eqs.(\ref{eq:C_3}-\ref{eq:C_4}) 
  allows us to derive an equation for the time evolution of $P_{||}^\alpha$ and  $ P_{\perp}^\alpha$ .  Noting that  $\Pi_{ij}^{(0, \alpha)}\partial(b_ib_j)/\partial t =0$, since $|{\bf b}|=1$ by definition,   we obtain  
  \begin{equation}\label{eq:C_5}
\frac{d P_{||}^\alpha}{d t ^\alpha}+P_{||}^\alpha\frac{\partial u_k^\alpha}{\partial x_k}=-2P_{||}^\alpha b_l b_k\frac{\partial u_l^\alpha}{\partial x_k},
\end{equation}
\begin{equation}\label{eq:C_6}
\frac{d P_\perp^\alpha}{d t ^\alpha}+2P_{\perp}^\alpha\frac{\partial u_k^\alpha}{\partial x_k}=P_{\perp}^\alpha b_l b_k\frac{\partial u_l^\alpha}{\partial x_k}.
\end{equation}
 Eqs.(\ref{eq:C_5}-\ref{eq:C_6}) represent the most general writing of the double adiabatic equations, as  deduced in a two fluid model. These reduce to the usual CGL closure \cite{CGL} of the single-fluid model when we can assume $|{\bm u}^e|\sim|{\bm u}^i|$. 
  Eqs.(\ref{eq:C_5}-\ref{eq:C_6}), extended to include $1^{st}$-order FLR corrections and the vectorial heat-flux contribution, were recently re-derived in Ref.\cite{Cerri_2013}, by ordering $\omega/\Omega_\alpha\sim 1/(\tau_H^\alpha\Omega^\alpha)\sim \rho_\alpha/L_H$ with $L_H^{-1}$ characteristic spatial gradient  of the fluid velocity.
 
\section{Small FLR limit for the magnetosonic waves at perpendicular propagation}\label{Appendix_C}
We order  $\omega/\Omega\sim kv_{th}/\Omega\sim\varepsilon \ll 1$.
We see from Eq.(\ref{eq:linear_ux}) and Eq.(\ref{eq:linear_tens_xx}) that the ordering $\tilde{u}_x/v_{th}\sim \tilde{\Pi}_{xx} {/(n_0m v_{th}^2)}\sim\varepsilon^0$ and $\tilde{\Pi}_{xy} {/(n_0m v_{th}^2)}\sim\varepsilon$ can be consistently assumed while from Eq.(\ref{eq:linear_tens_yy}) we can assume $\tilde{\Pi}_{yy} {/(n_0m v_{th}^2)}\sim\varepsilon^0$. However, we see from Eq.(\ref{eq:linear_uy}) that the ordering $\tilde{u}_y/v_{th}\sim\varepsilon$ follows, which is indeed coherent with the LF magnetosonic branch polarisation, for which  {$\tilde{u}_y/\tilde{u}_x\sim i\varepsilon$ (Eq.(\ref{eq:L_polarization_LFB}))}.  This latter condition implies through Eq.(\ref{eq:linear_tens_xy}) the ordering $(\tilde{\Pi}_{yy}-\tilde{\Pi}_{xx}) {/(n_0m v_{th}^2)}\sim\varepsilon^2$. We may 
now write 
\begin{equation}\label{eq:App_C_1}
\tilde{\Pi}_{yy}=\tilde{P}_\perp^0 + \tilde{\Pi}_{yy}^{(1)},\qquad 
\tilde{\Pi}_{xx}=\tilde{P}_\perp^0 - \tilde{\Pi}_{yy}^{(1)},
\end{equation}
where the apex ``$(1)$'' labels the $\sim\varepsilon^2$ correction. This is to ensure  the conservation of the trace of $\Pi_{ij}$ at any order in  $\varepsilon$, as suggested by the sum of Eq.(\ref{eq:linear_tens_xx}) and Eq.(\ref{eq:linear_tens_yy}).

{Using Eq.(\ref{eq:C_6}), that gives  $\tilde{P}_\perp^0=2P_\perp^0 k\tilde{u}_x/\omega$,} from Eq.(\ref{eq:linear_tens_xx}) we obtain (cf. Eq.(A17) of \cite{Cerri_2013})
\begin{equation}\label{eq:App_C_2}
\tilde{\Pi}_{xy}=i\frac{P_\perp^0}{2}\frac{k\tilde{u}_x}{\Omega}.
\end{equation}
Here, only the $\sim \varepsilon$ contribution has been retained, the other terms being  at least of order $\sim\varepsilon^3$.  On the contrary   the last term  in  Eq.(\ref{eq:linear_tens_xy})   leads to an  $\sim \varepsilon^2$ contribution:  combining {Eq.(\ref{eq:App_C_2})} with Eq.(\ref{eq:App_C_1}) we  obtain
\begin{equation}\label{eq:App_C_3}
\tilde{\Pi}_{xx}^{(1)}=-\tilde{\Pi}_{yy}^{(1)}=-i\frac{P_\perp^0}{2}\frac{k\tilde{u}_y}{\Omega}
-\frac{\omega}{\Omega}\frac{P_\perp^0}{4}\frac{k\tilde{u}_x}{\Omega}.
\end{equation}
This last equation differs from the usual CGL-FLR contribution  because of the last term (see  Eq.(A16) of \cite{Cerri_2013} for $\partial_y=0$).  This term  is neglected when the maximal ordering $\tilde{\Pi}_{yy}-\tilde{\Pi}_{xx}\sim\varepsilon $
 related to the assumption $\tilde{u}_y/v_{th}\sim\tilde{u}_x/v_{th}\sim \varepsilon^0$ is performed in Eq.(\ref{eq:linear_tens_xy}), as  is normally done in the CGL-FLR model. Solving the linear system of Eqs.(\ref{eq:linear_ux}-\ref{eq:linear_uy},\ref{eq:linear_tens_xy}-\ref{eq:linear_tens_yy}) by using Eqs.(\ref{eq:App_C_1}-\ref{eq:App_C_3})
leads to  the dispersion relation and group velocity of {the small FLR limit of the LFB} (Eqs.(\ref{eq:FLR_limit_omega}-\ref{eq:FLR_limit_vg})), 
which gives an opposite dispersive correction  with respect to {the CGL-FLR magnetosonic mode} (Eqs.(\ref{eq:CGL_2}-\ref{eq:CGL_3})), besides having a further contribution linear in $v_{th}^2\rho_i^2$.

{ Though Eqs.(\ref{eq:FLR_limit_omega}-\ref{eq:FLR_limit_vg}) are consistent with the appropriate limit of Eq.(\ref{eq:Comparison_1}) when the heat-flux contribution is neglected, this latter assumption in a full kinetic treatment implies the restriction to small values of  $\beta$. In this regard we recall  that  it has been  pointed out in Ref. \cite{Mikhailovskii} how to recover also in the fluid $\beta\sim 1$ regime the correct kinetic small-FLR limit of the magnetosonic dispersion relation.

 At $\beta\sim 1$ the  $\sim k^4v_{th}^4/ (\omega^2 \Omega^2)$  contributions to the small frequency dispersion relation should be retained. For consistency we must  retain also the $\sim k^4v_{th}^4/ (\omega^2 \Omega^2)$  
contribution to the  $n=0$ harmonic  in $K_{22}$ of  Eq.(\ref{eq:model_V_3}) arising from the next order term in the expansion of the Bessel function, which leads to Eqs.(\ref{eq:K_22_corrected}-\ref{eq:D_22_corrected}) and then to  the kinetic dispersion relation of the magnetosonic branch that includes the FLR correction to the sound term, too (Eq.\ref{eq:Comparison_2}.
 The kinetic dispersion relation (\ref{eq:Comparison_2}) differs from Eq.(\ref{eq:FLR_limit_omega}) because of the factor $5$ in the $\sim k^2v_{th}^2k^2\rho_i^2$ term. The discrepancy is due to the neglect of the gradient of the heat flux from Eq.(\ref{eq:M_3}), whose lowest order contribution to the linear system was shown in Ref.\cite{Mikhailovskii} to enter with a term $\partial_x\tilde{Q}_{xxy} \sim \varepsilon^2$  in Eq.(\ref{eq:linear_tens_xy}).
 This thermal flux contribution, first derived in Ref.\cite{Mikhailovskii_2} and more recently re-discussed in Ref.\cite{Smolyakov}, would correct the r.h.s. term of Eq.(\ref{eq:App_C_3}) with a further term
$ -ik{\tilde{Q}_{xxy}}/(2\Omega)$, which  modifies  Eq.(\ref{eq:App_C_3}) into 
\begin{equation}\label{eq:App_C_4}
\tilde{\Pi}_{xx}^{(1)}=-\tilde{\Pi}_{yy}^{(1)}=-i\frac{P_\perp^0}{2}\frac{k\tilde{u}_y}{\Omega}
-\frac{\omega}{\Omega}\frac{P_\perp^0}{4}\frac{k\tilde{u}_x}{\Omega}\left( 1  - 2  \,  k^2\rho_i^2 \frac{\Omega^2} {\omega^2}\right).
\end{equation}}
\

The above discussion is summarised by the  long  wave-length limit of the dispersion matrix (\ref{eq:L_fluid_3}) for the  magnetosonic branch, as obtained by using  both Eqs.(\ref{eq:App_C_3}-\ref{eq:App_C_4}),   
{\begin{equation}\label{eq:App_C_5}
[{\bf M}]=\left(
\begin{array}{cc}
\displaystyle{1-\frac{k^2}{\omega^2}
\left[c_a^2 + v_{th}^2\left( 1-\frac{k^2\rho_i^2}{4}
\right)+\frac{k^2\rho_i^2}{8}\right]}
 & \qquad 
\displaystyle{i k\rho_i\frac{kv_{th}}{\omega}} 
  \\ \\
-\displaystyle{i
k\rho_i\frac{kv_{th}}{\omega}}
&\qquad 1
\end{array}
\right).
\end{equation}}
 In the standard CGL-FLR approximation the $k\rho_i$ terms are inconsistently set to zero everywhere but in the off-diagonal components.  Including the $k^2\rho_i^2/8$ term in square brackets of the $M_{xx}$ element, which is due to the last r.h.s term of Eq.(\ref{eq:App_C_3}), makes it possible  instead to recover the correct  magnetosonic dispersion relation of Eq.(\ref{eq:Comparison_1}) in the small-$\beta$ limit. If we let $kd_i\sim k\rho_i$, i.e. $\beta\sim 1$, the lowest order contribution of the heat-flux gradient should  also be retained, and from the further contribution  of Eq.(\ref{eq:App_C_4}) to Eq.(\ref{eq:App_C_3})    the $-k^2\rho_i^2v_{th}^2/4$ correction in the $M_{xx}$ element follows. In this case  Eq.(\ref{eq:Comparison_2}) is recovered. 

We conclude by noting that, while the  assumption $\tilde{u}_y/v_{th}\sim \varepsilon^0$ is  incorrect to describe the propagation of magnetosonic waves in this specific geometry where ${\bm B}_0\cdot{\bm k}=0$, it  becomes legitimate when the condition ${\bm B}_0\cdot{\bm k}=0$ is relaxed or when compressionless equilibrium conditions  dependent e.g. simply on ${\bm u}=(0, u_y(x),0)$ are searched for, as  done in Ref.\cite{Cerri_2013}. In the first case,  a contribution proportional to $k_{||}c_a^2$ due to the $y$-component of the $\Omega{\bm J}\times{\bm b}/(ne)$ force of Eq.(\ref{eq:M_2}) would appear at r.h.s. of Eq.(\ref{eq:linear_uy}), which allows us in principle to order $u_y/v_{th}\sim u_x/v_{th}\sim \varepsilon^0$, at least as long as $k_{||}/k_\perp$ is not ordered small (i.e. of order $\varepsilon$ itself). In the second case, since $u_y(x)$ only enters in the incompressible non-gyrotropic equilibria devised in Ref.\cite{Cerri_2013},  no relative ordering is needed between $u_x$ and $u_y$ (and heat fluxes are null at equilibrium). We finally note  that corrections to the inconsistent CGL-FLR description of the propagation of fast magnetosonic waves have  also been  discussed in the different context of the  Landau-fluid models \cite{Goswami,Passot,Sulem}.

\end{document}